\documentstyle[prd,aps,floats]{revtex}
\begin{document}
\draft

%%%%%%%%%%%%%%%%%%%%%%%%%%%%%%%%%%%%%%%%%%%%%%%%%%%%%%%%%%%%%%%%%%%%%%
%
% Uncomment following four lines and one below for 2 column format
% and figure insertions.
%
\input epsf \renewcommand{\topfraction}{1} 
\twocolumn[\hsize\textwidth\columnwidth\hsize\csname
@twocolumnfalse\endcsname
%%%%%%%%%%%%%%%%%%%%%%%%%%%%%%%%%%%%%%%%%%%%%%%%%%%%%%%%%%%%%%%%%%%%%%
\title{How does the geodesic rule really work for global symmetry breaking first order phase transitions?}
\author{Antonio Ferrera \thanks{e-mail: aferrera@pinar2.csic.es}}
\address{Instituto de Matematica Aplicada y Fisica Fundamental,\\
Consejo Superior de Investigaciones Cientificas,\\
Serrano 121, Madrid}
\date{15/9/97}
\maketitle

\begin{abstract}
The chain of events usually understood to lead to the formation of
topological defects during phase transitions is known as the Kibble
mechanism. A central component of the mechanism is the so-called ``geodesic
rule''. Although in the Abelian Higgs model the validity of the geodesic
rule has been questioned recently, it is known to be valid on energetic
grounds for a global U(1) symmetry breaking transition. However, even for
these globally symmetric models no dynamical analisys of the rule has been
carried to this date, and some points as to how events proceed still remain
obscure. This paper tries to clarify the dynamics of the geodesic rule in
the context of a global U(1) model. With an appropriate ansatz for the field
modulus we find a family of analytical expressions, phase walls, that accounts for both geodesic and nongeodesic configurations. We then show how the latter ones are unstable and decay into the
former by nucleating pairs of defects. Finnally, we try to give a physical perspective of how the geodesic rule might really work in these transitions.
\end{abstract}


\preprint{\vbox{
\hbox{CAT-95/05}}}

% This is the other line to be uncommented for 2 column format
\vskip2pc]
%%%%%%%%%%%%%%%%%%%%%%%%%%%%%%%%%%%%%%%%%%%%%%%%%%%%%%%%%%%%%%%%%%%%%%

\section{Introduction}

According to the current models in particle physics, symmetry breaking phase
transitions are expected to have occurred in the early stages of the
evolution of the universe. There are two mechanisms by which these
transitions may have taken place: either by the formation of bubbles of the
new phase within the old one (i.e., first order phase transition), or by
spinodal decomposition (i.e., second order transition). Although we still do
not know which one did actually take place in each particular case, for the
electroweak phase transition for instance common opinion inclines more
towards the first of these two possibilities. As the first order transition
scenario would have it, bubbles of the new phase nucleated within the old
one (the nucleation process being described by instanton methods as far as
the WKB approximation remains valid \cite{c77}), and subsequently expanded
and collided with each other until they occupied all of the available volume
at the time at which the transition was completed. In the process of bubble
collision though, the possibility arises that regions of the old phase
become trapped within the new one, giving birth to topologically stable
localized energy concentrations known as {\it topological defects } (for
recent reviews see refs. \cite{td-rev}), much in the same way in which these
structures are known to appear in condensed matter phase transitions.

From a theoretical point of view, topological defects will appear whenever a
symmetry group $G$ is spontaneously broken to a smaller group $H$ such that
the resulting vacuum manifold $M=G/H$ has a non-trivial topology: cosmic
strings for instance (vortices in two space dimensions) will form whenever
the first homotopy group of $M$ is non-trivial, i.e., $\pi _1(M)\neq 1.$ To
see how this could happen in detail, let us consider the Lagrangian

\begin{equation}
{\cal L}=\partial _\mu \Phi \partial ^\mu \Phi ^{*}-V(\Phi )  \label{lag}
\end{equation}
for a complex scalar field $\Phi $. Let us assume that $V$ is of the type $V=%
\frac \lambda 4(|\Phi |^2-\eta ^2)^2$, and that its parameters are functions
of the temperature such that at high temperatures $\Phi =0$ is the only
minimum of $V$, while at zero temperature all the $|\Phi |=\eta $ states
correspond to different degenerate minima. Then the structure of the vacuum
manifold will be that of $S^1$. $\pi _1(S^1)\neq 1$ however, and thus we can
form non-contractible loops in the vacuum manifold. That is, the model
admits cosmic string solutions.

The way in which these configurations would actually arise during a phase
transition is via the Kibble mechanism \cite{k76}. In the context of a first
order transition the basic ideas behind it are that bubbles are nucleated
with random phases of the field, and that when two regions in which the
phase takes different values encounter each other the phase should
interpolate between this two regions following a geodesic path in the vacuum
manifold --the so-called {\em geodesic rule}. A possible scenario for vortex
formation in two space dimensions would then look like this: three bubbles
with respective phases of $0,$ $2\pi /3,$ and $4\pi /3$ collide
simultaneously. Then, if we walk from the first bubble to the second one the
geodesic rule tells us that the phase has to grow from $0$ to $2\pi /3$
along our path (as opposed to decreasing from $0$ to $-4\pi /3$). In the
same manner, walking from the second bubble to the third one we will see the
phase grow from $2\pi /3$ to $4\pi /3$, and if we finally walk from the
third bubble back to the first one again we will see a change from $4\pi /3$
to $2\pi $. The phase will have thus wounded up by $2\pi $ in our path,
having traversed the whole of the vacuum manifold once along the way.
Continuity of the field everywhere in the region inside our path demands
then that the field be zero at some point in this region, namely the vortex
core. In the limit in which the bubbles extend to infinity, outwards from
the center of collision, removal of this vortex would cost us an infinite
amount of energy, since it would involve unwinding the field configuration
over an infinite volume. The vortex is thus said to be topologically stable.
In three space dimensions, the resulting object would obviously be a string,
rather than a vortex. Clearly though there are other ways in which strings
could be formed. Collisions of more than three bubbles could also lead to
string formation, or, for instance, two of the bubbles could collide first,
the third one hitting them only at some later time while the phase is still
equilibrating within the other two. This event in particular will be far
more likely than a simultaneous three way (or higher order) collision, and
it is probably the dominating process by which strings are formed
(especially if nucleation probabilities are as low as required for WKB
methods to be valid). When two bubbles collide, the phase will try to reach
a homogeneous distribution within the single true vacuum cavity formed after
the collision. However, in the absence of coupling between the scalar and
the plasma that would also be present in a cosmological setting, this
process is never completed --essentially because the velocity at which the
phase propagates inside the bubbles is the same as that with which the
bubble walls expand. Thus, this does not substantially modify the picture of
defect formation described above.

The geodesic rule is therefore central in that not only gives us a heuristic
view of how bubbles interact after they collide, but it also yields an
estimate of the density of defects produced in a phase transition, i.e.,
about one per every three bubbles (times the probability that the three
bubbles have the correct phases of course). If one is concerned about the
density of defects that are present after a cosmological phase transition,
the question of the validity of the rule is therefore of obvious importance.
Despite its importance however, the rule still presents serious problems in
its understanding. For globally symmetric models the role of the geodesic
rule was studied by Srivastava some time ago. In \cite{s92} he established
its validity within these type of models mainly on energetic grounds, the
geodesic path being the least energetic of all the possible paths that the
phase can follow. Recently however a number of authors have argued for and
against the validity of the rule \cite{geodesic}, or observed what seems to
be a breakdown of the rule in numerical simulations \cite{CS}, within the
context of the Abelian Higgs model. In this case it is rather less clear
whether one can reason on energetic grounds since the phase difference
between two points is not a gauge invariant quantity by itself. In principle
therefore one could always choose a gauge in which the phase is zero
everywhere, and it would not be immediately obvious how the geodesic rule
would work in this situation. The goal of this paper is to take a step back
and try to give a dynamical account of how the geodesic rule may actually
work for a globally U(1) symmetric model. In order to do this we will use a
type of analytical solutions, hereafter referred to as {\it phase walls,} by
which the phase interpolates between its values within two bubbles. We will
also resort to numerical simulations to study the stability of different
phase wall configurations, and finally we will discuss the physical settings
in which non-geodesic phase walls can appear. The paper is organized as
follows: in section II we give a brief account of the model and the results
previously found in \cite{fm96} relevant for this paper. We then extend our
treatment in section III to include the case in which the scalar field is
not dissipatively coupled and we have 2+1 dimensions. In section IV we study
the stability of non-geodesic phase walls. In V we study the influence of
our work in three bubble collisions and topological defect nucleation, and
in VI we present a general view of how the geodesic rule might work in
globally symmetric models. Finally, the conclusions and the implications
that our analysis may have for the Abelian Higgs mode are presented in
section VII.

\section{1 Dimensional phase walls in a damping environment}

\subsection{expression for the interpolating phase}

Consider the Lagrangian (\ref{lag}) for a complex field $\Phi $. We will use
the same form of potential that was used in \cite{s92,mp95}, that is, 
\begin{equation}
V=\lambda \left[ \frac{|\Phi |^2}2\left( |\Phi |-\eta \right) ^2-\frac %
\epsilon 3\eta |\Phi |^3\right] .  \label{pot}
\end{equation}
This is just a quartic potential with a minimum at $|\Phi |=0$ (the false
vacuum), and a set of minima connected by a $U(1)$ transformation (true
vacuum) at $|\Phi |=\rho _{tv}\equiv \frac \eta 4(3+\epsilon +\sqrt{\left(
3+\epsilon \right) ^2-8})$, towards which the false vacuum will decay via
bubble nucleation. It is the dimensionless parameter $\epsilon $ that is
responsible for breaking the degeneracy between the true and the false vacua.

The equations of motion for this system are then 
\begin{equation}
\partial _\mu \partial ^\mu \Phi =-\partial V/\partial \Phi .  \label{em}
\end{equation}
For the potential (\ref{pot}), approximate solutions of (\ref{em}) exist for
small values of $\epsilon$, the so-called thin wall regime \cite{l83}, and
are of the form

\begin{equation}
|\Phi |=\frac{\rho _{tv}}{2}\left[ 1-\tanh \left( \frac{\sqrt{\lambda }\eta 
}{2}(\chi -R_{0})\right) \right] ,  \label{thin}
\end{equation}
where $R_{0}$ is the bubble radius at nucleation time and $\chi ^{2}=|%
\stackrel{\rightarrow }{x}|^{2}-t^{2}$. The bubble then grows with
increasingly fast speed and its walls quickly reach velocities of order 1.
In \cite{fm96} however the authors were concerned about a model with
overdamped motion of the walls due to the interaction with a surrounding
plasma. In order to model this effect, they inserted a frictional term for
the modulus of the field in the equation of motion, namely 
\begin{equation}
\partial _{\mu }\partial ^{\mu }\Phi +\gamma \stackrel{\cdot }{|\Phi |}%
e^{i\theta }=-\partial V/\partial \Phi ,  \label{damp-em}
\end{equation}
where $\stackrel{\cdot }{|\Phi |}\equiv $ $\partial |\Phi |/\partial
t,\theta $ is the phase of the field, and $\gamma $ stands for the friction
coefficient --which will serve as parameter under which we will hide our
lack of knowledge about the detailed interaction between the wall and the
plasma. It can be shown (see \cite{fm96}) that this equation does indeed
posses a solution that shows the desired type of overdamped motion for the
bubble walls. In the thin wall limit, this solution can be written as

\begin{equation}
\rho =\frac{\rho _{tv}}{2}\left[ 1-\tanh \left( \frac{\sqrt{\lambda }\eta }{2%
}\frac{(r-v_{ter}t-R_{0})}{\sqrt{1-v_{ter}^{2}}}\right) \right] ,
\label{wall}
\end{equation}
which is simply a spherically symmetric Lorentz-contracted moving domain
wall with a velocity $v_{ter}$ of the form $v_{ter}\sim \epsilon \delta
_{m}/\gamma \rho _{tv}^{2}$, where $\delta _{m}=1/(\sqrt{\lambda }\eta )$ is
the bubble wall thickness.

In two bubble collisions, the phase will at first --well before the bubble
walls actually collide, at least in a zero temperature situation --
interpolate between the values that it takes in each bubble by means of a
phase wall situated at the midpoint in between the bubbles. This can be seen
by taking the value of the modulus of the field $\rho $ far away from the
bubbles as that given by the ansatz

\begin{eqnarray}
\Phi (bubble1+bubble2) \equiv \rho e^{i\theta }\simeq   \nonumber \\
\Phi (bubble1)+\Phi (bubble2)& \equiv \rho _{1}e^{i\theta _{1}}+\rho
_{2}e^{i\theta _{2}}.  \label{apsol}
\end{eqnarray}
Note that we will assume that (\ref{apsol}) yields only a correct
approximation for $\rho $, and not for $\theta $. For $\rho _{1,}\rho _{2}$
we will take an exponentially decaying ansatz in a 1 dimensional
approximation

\begin{eqnarray}
\rho _{1} &\simeq &\rho _{tv}e^{(x+vt-x_{0})/\delta _{m}},  \label{asympt} \\
\rho _{2} &\simeq &\rho _{tv}e^{-(x-vt+x_{0})/\delta _{m}},  \nonumber
\end{eqnarray}
where $v$ is the terminal velocity of the walls, $v_{ter}$ above, and the
bubble centers are situated at $\pm x_{0}$, with $R\lesssim x_{0},\,R\gg |$ $%
x_{0}-R|\gg \delta _{m}$ ($R$ being the bubbles radii). That is, we are
basically asking that the distance from the bubble walls to the origin $|$ $%
x_{0}-R|$ be several times $\delta _{m}$ (so that we fields decay
exponentially there), and to have thin wall bubbles with $R$ large enough so
that the 1-d approximation is valid. Then (\ref{apsol}) yields for the
combined modulus 
\begin{equation}
\rho ^{2}\simeq \rho _{tv}^{2}e^{2(vt-x_{0})/\delta _{m}}\left\{ 2\cosh
(2x/\delta _{m})+2\cos (\theta _{1}-\theta _{2})\right\} .  \label{cmodulus}
\end{equation}
We can now use this expression in the equation of motion for the phase $%
\theta $, to get 
\begin{equation}
\stackrel{..}{\theta }-\ \theta ^{\prime \prime }+\frac{2v}{\delta _{m}}%
\stackrel{.}{\theta }-\ \frac{2\delta _{m}^{-1}\sinh (2x/\delta _{m})}{\cosh
(2x/\delta _{m})+\cos (\theta _{1}-\theta _{2})}\theta ^{\prime }=0,
\label{phasemotion}
\end{equation}
where $\theta ^{\prime }\equiv \partial \theta /\partial x,\stackrel{.}{%
\theta }\equiv \partial \theta /\partial t.$ Note that for an initial phase
difference $\theta _{1}-\theta _{2}=\pi $ the modulus of the field (\ref
{cmodulus}) is zero at the midpoint between the bubbles. In this case then
the denominator in the last term of (\ref{phasemotion}) goes to zero due to
the fact that the phase has the shape of a step function as we go over the
origin, switching discontinuously from $\theta _{1}$ to $\theta _{2}$. On
the other extreme, if the initial phase difference is zero there is of
course no dynamics to it. We will then focus in an intermediate situation,
and find solutions to (\ref{phasemotion}) for a phase difference of $\pi /2$%
. Using the ansatz $\theta =T(t)X(x)$, we can separate variables to obtain: 
\begin{mathletters}
\begin{eqnarray}
\stackrel{..}{T}+\ \frac{2v}{\delta _{m}}\stackrel{.}{T}+\ k^{2}T &=&0,
\label{phasetime} \\
X^{\prime \prime }+\frac{2}{\delta _{m}}\tanh (\frac{2x}{\delta _{m}}%
)X^{\prime }+k^{2}X &=&0,  \label{phasex}
\end{eqnarray}
$k$ being the separation constant. The equation for $T$ is obviously that of
a damped oscillator. For $k=0$, a $T=const.$ solution exists consistent with
our boundary condition that $\theta $ goes to $\theta _{1,2}$ as $%
x\rightarrow \mp \infty $ at all times, while, for $k\neq 0$ all solutions
will present unwanted temporal dependence temporal dependence of $\theta $
even as $x\rightarrow \mp \infty $. (Note that in case the lifetime of any
transient configuration that might appear is only of the order $\delta _{m}/v
$, since the equation for $T$ is that of a damped oscillator). We take $k=0$
then, whence an immediate integration of (\ref{phasex}) leads to the
equation for $X^{\prime }$ (or equivalently $\theta ^{\prime }$ since $%
T=const.$) 
\end{mathletters}
\begin{equation}
\theta ^{\prime }=\frac{K}{\cosh \left( 2x/\delta _{m}\right) },
\label{firderiv}
\end{equation}
with $K$ being an integration constant with dimensions (length)$^{-1}$. This
equation admits a family of general solutions given by the static
sine-Gordon kink form 
\begin{eqnarray}
\theta & =K\delta _{m}\arctan (\exp (2x/\delta _{m}))+D=  \nonumber \\
& C\arctan (\exp (2x/\delta _{m}))+D,  \label{pwall2}
\end{eqnarray}
with $K\delta _{m}=C$, as can be easily checked by direct differentiation.
(Note that this is not the original form in which these solutions appeared
in \cite{fm96}, the expression found there is equivalent to this standard
sine-Gordon form though). The solution to the equation of motion for the
phase (\ref{phasemotion}) interpolating from say $\theta _{1}=0$ at $%
x\rightarrow -\infty $ to $\theta _{2}=\pi /2$ at $x\rightarrow +\infty $
will then simply be

\begin{equation}
\theta \left( x\right) =\arctan (\exp (\frac{2x}{\delta _{m}})),
\label{SGpwall}
\end{equation}
which clearly shows the structure of a phase wall placed at the origin and
of width closely related to the bubble wall's width $\delta _{m}$ --although
this relation will change as the phase difference between the bubbles in (%
\ref{phasemotion}) change. It is also clear from this expression that this
wall interpolates from $\theta _{1}$ to $\theta _{2}$ following the shortest
(geodesic) path in the vacuum manifold, since $\theta $ monotonically
increases from $0$ to $\pi /2$.

Since $\theta $ admits sine-Gordon solutions though, it must be possible to
write its equation of motion (\ref{firderiv}) in a sine-Gordon manner. Let
us use the form of $\theta (x)$ given by (\ref{pwall2}) and assume without
loss of generality that the phase at $x\rightarrow -\infty \,$is zero, so
that $D=0$. It is then not difficult to see that 
\begin{equation}
\sqrt{\frac{C^{2}}{2\delta _{m}^{2}}\left[ 1-\cos 4\left( \frac{\theta
\left( x\right) }{C}\right) \right] }=\ \frac{C}{\delta _{m}}\frac{1}{\cosh
\left( 2x/\delta _{m}\right) }.  \label{presineGordon}
\end{equation}
Therefore, with $V(\theta )=\frac{C^{2}}{4\delta _{m}^{2}}\left[ 1-\cos
\left( 4\theta /C\right) \right] ,$ the expression 
\begin{equation}
\frac{d\theta }{dx}=\sqrt{2V(\theta )}  \label{sineGordon}
\end{equation}
is a sine-Gordon equation which has (\ref{pwall2}) as its general solution. $%
V$ has its minima located at $\theta =mC\pi /2$, $m$ being an integer. Note
though that (\ref{sineGordon}) is a static sine-Gordon equation, thus it
only contains the kink and antikink sectors (apart from the vacuum of
course) and its solutions only connect adjacent vacua.

Going back to our particular problem, where $\Delta \theta =\theta
_{2}-\theta _{1}=\pi /2$, we can now use the fact that the physical phase $%
\theta $ is defined only up to modulo $2\pi $ to choose $\theta _{2}$ to be $%
\theta _{2}=\pi /2+2\pi n$, with $n=...-2,-1,0,1,2,....$ being an integer.
It immediately follows that (\ref{SGpwall}) is not the only solution
consistent with our physical boundary conditions. In fact there is an
infinite family of solutions to (\ref{firderiv}) having $\Delta \theta =\pi
/2+2\pi n$ which will satisfy our requirements, of which (\ref{SGpwall}) is
only the $n=0$ case. An interesting case corresponds to $n=-1$. This phase
wall forces $\theta $ to monotonically decrease from $\theta _{1}=0$ at $%
x\rightarrow -\infty $ to $\theta _{2}=-3\pi /2$ at $x\rightarrow +\infty ,$
and therefore follows the shortest non-geodesic path in the vacuum manifold
between $\theta _{1}$ and $\theta _{2}$. Its form is 
\begin{equation}
\theta \left( x\right) =-3\arctan (\exp (\frac{2x}{\delta _{m}})).
\label{nongeo}
\end{equation}
Lower negative values of $n$ will correspond to phase walls that completely
wind around the vacuum manifold $-|n|+1$ times in the negative $\theta $
direction before going from $0$ to $-3\pi /2$, whereas positive values of $n$
will lead to phase walls carrying a (positive) winding number $n$.
(Hereafter in the paper, when using the word ``winding'' to denote the phase
change across a phase wall it should always be understood that there is no
phase circulation around any closed path in physical space, as opposed to
the times in which ``winding'' is used to refer to the circulation of the
phase around a vortex or antivortex). The expression giving $\theta (x)$ for
a general value of $n$ will then simply be 
\begin{equation}
\theta \left( x\right) =\left[ 1+4n\right] \arctan (\exp (\frac{2x}{\delta
_{m}})).  \label{nongeogen}
\end{equation}

\section{Phase walls in 2 spatial dimensions without dissipation}

We will now extend the analysis carried out previously to the case in which
there is no dissipative coupling of the scalar field to the rest of the
matter present in the model. Also, and since we will be needing it in order
to be accurate in the numerical simulations that we will perform, we will
introduce 2 spatial dimensions in the problem.

The problem of two bubbles approaching each other and colliding in 2+1
dimensions is invariant under the SO(1,1) Lorentz group. If we take the
bubbles to nucleate initially at (0,$\pm x_{0}$,0) with coordinates (t,x,y)
and signature ($-,$+,+), then the symmetry of the problem dictates that we
must have $\rho =\rho (x,\tau )$ and $\theta =\theta (x,\tau )$, where $\tau
^{2}=y^{2}-t^{2}$. Because of the particular expression that we will use as
the approximate value of the modulus of the field though, a different set of
coordinates turn out to be more convenient. We define new coordinates $%
u,v,\alpha $ 
\begin{mathletters}
\begin{eqnarray}
x &=&x_{0}\cosh u\cos v, \\
y &=&x_{0}\sinh u\sin v\cosh \alpha , \\
t &=&x_{0}\sinh u\sin v\sinh \alpha .
\end{eqnarray}
Note that 
\end{mathletters}
\begin{equation}
\tau =x_{0}\sinh u\sin v,  \label{tau}
\end{equation}
thus to preserve Lorentz invariance in these coordinates is simply to say
that $\rho $ and $\theta $ can not depend on $\alpha $. The range of
variation of $u,v,\alpha $ will be 
\begin{equation}
u\geq 0;\qquad -\pi /2\leq v\leq \pi /2;\qquad \alpha \geq 0,  \label{range}
\end{equation}
and in terms of these coordinates the equation of motion for $\theta $ will
read (dropping already any dependence on $\alpha $) 
\begin{eqnarray}
\partial _{uu}\theta +\partial _{vv}\theta +\coth u\cdot \partial _{u}\theta
+\cot v\cdot \partial _{v}\theta \nonumber \\
+\frac{2}{\rho }\left( \partial _{u}\rho
\partial _{u}\theta +\partial _{v}\rho \partial _{v}\theta \right) =0.
\label{phamo1}
\end{eqnarray}
Now we have to specify an ansatz for $\rho $. As in the case before, we will
take the modulus of the field between the bubbles to be given by the modulus
of the sum of the two individual bubble fields. Thus, 
\begin{equation}
|\Phi |^{2}=\rho ^{2}=|\rho _{1}e^{i\theta _{1}}+\rho _{2}e^{i\theta
_{2}}|^{2}=\rho _{1}^{2}+\rho _{2}^{2}+2\rho _{1}\rho _{2}\cos (\Delta
\theta ).  \label{ro}
\end{equation}
Let us now define the invariant distance of any point to the bubble
nucleation events, $\chi _{1,2}$%
\begin{equation}
\chi _{1,2}=\sqrt{\left( x\pm x_{0}\right) ^{2}+y^{2}-t^{2}},
\label{distance}
\end{equation}
where the plus sign holds for $\chi _{1}$and the minus for $\chi _{2}$.
Looking back at (\ref{thin}) it is obvious that outside the bubbles and away
of the their walls we should have 
\begin{equation}
\rho _{1,2}\simeq \rho _{tv}e^{-\left( \chi _{1,2}-R_{0}\right) /\delta
_{m}}.  \label{approxrad}
\end{equation}
Taking this into (\ref{ro}), setting already $\Delta \theta =\pi /2$, and
writing the result in terms of $u$ and $v$ leads to an expression for $\rho $
\begin{equation}
\rho ^{2}=2\rho _{tv}^{2}e^{2R_{0}/\delta _{m}}e^{-\frac{2x_{0}}{\delta _{m}}%
\cosh u}\cosh \left( \frac{2x_{0}}{\delta _{m}}\cos v\right) ,  \label{ro2}
\end{equation}
as can be checked by trivial, if lengthy, manipulations. Then plugging this
form for $\rho $ into (\ref{phamo1}) results in 
\begin{eqnarray}
\partial _{uu}\theta +\partial _{vv}\theta +\left[ \coth u-\frac{2x_{0}}{%
\delta _{m}}\sinh u\right] \partial _{u}\theta +  \nonumber \\
\left[ \cot v-\frac{2x_{0}}{\delta _{m}}\sin v\tanh \left( \frac{2x_{0}}{%
\delta _{m}}\cos v\right) \right] \partial _{v}\theta =0  \label{phamo2}
\end{eqnarray}
In spite of the appearance of this equation, it is actually not too
difficult to find the type of solutions that we are looking for, at least
within some approximation. Note that we expect the phase walls to have their
central axis located at $x=0$, the line equidistant to the bubble centers,
which in $u,v$ coordinates corresponds to $v=\pi /2$. If we expand the
second bracket in powers of $v$ around $v_{0}=\pi /2$ it is easy to see that
neglecting the $\cot v$ term when compared to the other term in the bracket
amounts to neglecting terms of order $1$ or smaller when compared to terms
of order $\left( 2x_{0}/\delta _{m}\right) ^{2}$ or higher in the
coefficients of the series. Since within the approximation that we are
working with $x_{0}/\delta _{m}\gg 1$, we conclude that to an accuracy $%
1/\left( 2x_{0}/\delta _{m}\right) ^{2}$ we can safely neglect the first
term in that bracket. Assuming as ansatz that the solutions that we are
looking for depend only in $v$ leads then to 
\begin{equation}
\partial _{vv}\theta -\frac{2x_{0}}{\delta _{m}}\sin v\tanh \left( \frac{%
2x_{0}}{\delta _{m}}\cos v\right) \partial _{v}\theta =0  \label{phamo3}
\end{equation}
which indeed has phase wall solutions of the form 
\begin{equation}
\theta =\left[ 4n+1\right] \arctan (\exp (\frac{2x_{0}}{\delta _{m}}\cos v)),
\label{phawall}
\end{equation}
or, in terms of $\chi _{1,2}$%
\begin{equation}
\theta =\left[ 4n+1\right] \arctan (\exp (\frac{\chi _{1}-\chi _{2}}{\delta
_{m}})).  \label{2dphawall}
\end{equation}
Expression (\ref{2dphawall}) is clearly again a sine-Gordon kink, differing
from (\ref{nongeogen}) only in its argument. Note however that the fact that
(\ref{2dphawall}) depends on $\chi _{1}-$ $\chi _{2}$ implies that it does
not represent a planar wall. This is depicted in figure \ref{2dpwall}, where
we show constant phase lines for the geodesic phase wall interpolating from $%
0$ to $\pi /2$. Thus, lines of constant phase will have a curvature which
will depend on the distance to the bubble centers (in the figure the bubble
centers are situated at $x_{0}=\pm 10$). If we expand $\chi _{1,2}$ in terms
of $\overrightarrow{\chi }/x_{0}$ though, where $\overrightarrow{\chi }%
=\left( t,x,y\right) $, (\ref{2dphawall}) reduces to the 1-dimensional
expression (\ref{nongeogen}) above for as long as $\left( x/x_{0}\right)
^{2}\sim \left( y/x_{0}\right) ^{2}\sim \left( t/x_{0}\right) ^{2}\ll 1$. As
in the case studied in the previous section, $n=0$ will correspond to the
geodesic phase wall solution, while higher (or lower) values of $n$ lead to
phase walls that wind around the vacuum manifold of the model $n$ (or $|n|-1$%
) times. As before, $n=-1$ corresponds to the shortest non-geodesic path. We
have thus shown that the existence of phase wall solutions is a general
feature of the model.

\section{Phase wall stability analysis}

Since an analytical approach proved to be too involved, we resorted to
numerical simulations in order to investigate the problem of the stability
of non-geodesic phase walls --the geodesic one being obviously stable. We
used a straight forward leapfrog method with second order accuracy to evolve
in time different initial configurations corresponding to phase walls of
various winding numbers. In order to better understand the process of phase
wall decay we first studied the evolution of the 1-dimensional solutions,
then carried on the study to two dimensional configurations.

We begin in figure \ref{1dnm1} by plotting, using polar coordinates in field
space, the fate of the shortest nongeodesic wall through a series of
snapshots. In \ref{1dnm1}(a) we can see the initial configuration of the
phase at $\ t=0$, going from $\theta =0$ to $\theta =-3\pi /2$. Subsequently
we see how in the very core of the wall, the field starts decreasing its
modulus, eventually crossing the origin ($\rho =0$), and finally settling
down in the geodesic configuration, in which $\theta $ goes from $\theta =0$
to $\theta =\pi /2$, by $t=2$ in units of the bubble wall thickness $\delta
_{m}$. In figure \ref{1dnm2} we have depicted the unwinding of a $n=-2$
wall. In \ref{1dnm2}(a) we can see how the initial configuration has, as
expected, an extra $2\pi $ winding over the $n=-1$ wall previously seen.
Note how in this case the decay proceeds by unwinding first this extra $2\pi 
$ loop, after which we are left with a configuration equivalent to that of a 
$n=-1$ wall which in turn decays as described above. The unwinding of the
extra $2\pi $ is shown in figure \ref{1dnm2b}. In it we can see how the loop
is progressively displaced and shrunk, until the field crosses the origin at
the midpoint between the bubble centers. After that, the field reeqnarrays
itself. The gradient energy acts as a tension on the phase loop, which
eventually evolves into a cusp. This cusp then divides into two sharp kinks
that move away from each other. As we can see then the main features that
seem to emerge of these results are that a phase wall of winding $n$ will
cross the zero field value $|n|$ times in its decay, decreasing $n$ in steps
of one, and that the time scale of the decay is of the order of $\delta _{m}$%
, becoming progressively shorter as the winding increases.

Using these results it is not too difficult to convince oneself of the fact
that in two spatial dimensions we should in general expect the decay process
to produce pairs of vortices and antivortices. Let us think for instance of
the symmetry axis of a $n=-1$ phase wall, that is, the line of points that
are equidistant to the bubble centers. (For a 1-dimensional wall, the
symmetry axis is of course reduced to a single point: the central point
equidistant to the bubble centers). In two space dimensions we may think of
the $n=-1$ wall\ decay as being the result of all the points in this axis
crossing the zero field value --each following a similar process as the
1-dimensional case-- and reeqnarraying to follow the geodesic path. The points
along the wall's symmetry axis are not all equivalent to one another though,
since the distance to the bubble centers changes as we move along the axis.
Therefore the field does not cross the zero value everywhere along the axis
simultaneously, but rather some points will go through this value before
others. In particular, given the symmetry of the problem it seems clear that
the central point of the axis, with the smallest distance to the bubble
centers, will cross the zero first. Once the field has switched to the
geodesic path along the line that joins the bubble centers, the two points
immediately above and below the central point will cross the zero
themselves, then the ones respectively above and below these two, and so on.
Therefore, as the phase shifts towards the geodesic path we always have a
pair of points on the wall axis at which the field takes the false vacuum
value, both of them quickly moving up and down respectively. Since the
direction of the field in the region that has already switched to the
geodesic path is opposite to its direction immediately above and below it,
we end up having a net circulation of the phase around these two points. We
have thus produced a vortex-antivortex pair along the phase wall axis, the
two defects moving away from each other.

In figure \ref{2dnm1} we can see how this process takes place. In it we can
see the contour lines of the modulus of the field, while the phase is
represented by the angle that the arrows form with the $x$ axis. The length
of the arrows is proportional to the modulus of the field at the point at
which it is drawn, except in the central region with $\rho <0.001$ where
they have been magnified to a common length so that we can appreciate the
decay process. Throughout the simulations the true vacuum value of the field
is $\rho _{tv}\sim 1$. In figure \ref{2dnm1}(a) we have the initial $n=-1$
configuration, if we follow the change of the phase horizontally along the
central $y=0$ line for instance we can see how the phase steadily decreases
from $0$ to $-3\pi /2$. In figure \ref{2dnm1}(b) we see how the field starts
decreasing its modulus, especially in the central region between the two
bubbles. Eventually, as described above, the central point will cross the $%
\rho =0$ value first. After this has occurred the points above and below it
will follow suit, then the ones respectively above and below these two, and
so on. Due to the opposing phase directions above and below the $\rho =0$
points though a vortex-antivortex pair will form, as can be seen in \ref
{2dnm1}(c). As successive points along the axis cross the zero, the defects
move further and further away from each other, as shown in \ref{2dnm1}(d).
Note how on \ref{2dnm1}(d) the phase still has some reeqnarrayment left to do
in the region that has crossed the zero field value. It is below zero right
before the wall axis, and over $\pi /2$ right after it as we move from left
to right. This situation is analogous to what we see in figure \ref{1dnm1}%
(b) and (c) in the 1-dimensional case. Eventually, the geodesic
configuration is attained.

In figure \ref{2dnm2} we can see the same process taking place for the $n=-2$
phase wall. As was the case with the 1-dimensional solution, this
configuration first decays by unwinding its extra $2\pi $ to become a $n=-1$
wall. In order to do this however it must cross the $\rho =0$ value as we
saw in 1 dimensions. The $2\pi $ unwinding will thus make the field cross
its zero value, first at the origin, then generate two zeroes of the field
that move away from each other along the wall axis following a process
analogous to the one just seen. As before thus, a pair of defects is created
in this decay. Although not shown in the figures, the $n=-1$ wall will then
in turn decay to the geodesic configuration generating a second pair of
defects. Therefore, the final unwinding of the $n=-2$ wall produces two
pairs of defects. The rule that immediately seems to suggest itself is that
a phase wall of winding $n$ will produce precisely $|n|$ pairs of defects in
its decay. The simple reason behind it would be that for such a wall the
field needs to cross $|n|$ times the $\rho =0$ value in order to complete
its decay, and that each time that the field crosses the zero value at the
central point a vortex-antivortex pair is produced. One should be careful
though. Note that the decay of the $n=-2$ wall to the $n=-1$ configuration
sends and vortex into the lower half axis of the wall, whereas the decay of
the $n=-1$ wall sends an antivortex into the same region. Since global
vortices and antivortices attract each other one could worry that these
pairs could tend to annihilate, especially in the case of high $n$ walls
where the defects will be produced within a short distance of each other. It
would seem though that such an annihilation can not actually happen, since
each defect carries away as phase circulation around it one of the winding
factors present in the wall. (I.e., think for instance of the $n=-2$ case.
The successive vortex and antivortex produced in the decay can not
annihilate each other because, if they did, the phase would have to go from
the region in which the wall has not decayed yet into the region in which it
has, having no zeros around which it can wind in between). It is unclear at
the moment whether this same argument holds for higher $n$ walls however.
More simulations should perhaps be carried to test this case.

\section{Non geodesic phase walls and three bubble collisions: topological
defect nucleation}

If non geodesic phase walls can indeed be formed in two bubble collisions
and they subsequently decay by nucleating vortices, then the next obvious
question is: what will be the impact of these processes on the probability
of topological defect becoming trapped in a region of true vacuum when three
bubbles collide?. The standard picture of defect nucleation described in the
introduction roughly yields a prediction of $1$ defect per every three
bubbles (times a proportionality constant to account for the probability of
the three bubbles having the appropriate phase). Let us think of three
colliding bubbles with phases such that they would nucleate a defect
following the geodesic rule: is the defect still formed when the phase
chooses a non-geodesic path between two of the bubbles?. Alternatively, if
the initial phases would not yield a vortex following the geodesic rule,
will a vortex be nucleated if a non-geodesic wall is formed between any two
bubbles?. Clearly, these questions can potentially alter the standard
prediction for vortex nucleation probability. We have therefore simulated
these two scenarios to find out their answers.

In figure \ref{3bvortex} we can see the results of the first case. In it we
have three bubbles with phases $\pi /6$, $5\pi /6$ and $3\pi /2$ that in
principle would nucleate a vortex following the geodesic rule. The two upper
bubbles however have collided slightly before, hitting the third one a
little later, and the phase has wound up between them following a $n=-1$
phase wall. As we can see in figure \ref{3bvortex}(a) though, the field does
not reach its zero value anywhere in the region between the three bubble
centers, as it would be the case if the phase were to follow the geodesic
path everywhere. This is the key of the question here: if the geodesic rule
were to be holding, the vortex predicted by it would appear in the region
between the three bubbles as soon as they make contact with each other, in
this case however the vortex is not there to begin with. As we have seen
though, the phase wall has to decay producing a vortex-antivortex pair. In
the following figures we can then see how the vortex produced in the decay
ends up resting between the three bubbles, while the antivortex escapes
``upwards'' and away from the bubbles, into a region that gets exponentially
closer and closer to the false vacuum. Thus, in terms of stable vortices
--that is, vortices surrounded by regions of true vacuum-- the net result is
exactly the same as the one predicted by the geodesic rule. There is
however, one extra antivortex that has been expelled into the near-false
vacuum region.

The second case is depicted in figure \ref{3bnovortex}. Clearly the phases
inside the bubbles would not produce a vortex should the geodesic rule hold.
As before, a $n=-1$ phase wall has formed between the upper two bubbles, and
in this case the field already has a zero in the region between the bubble
centers as we can see in figure \ref{3bnovortex}(a). When the phase wall
unwinds, the resulting vortex annihilates with the existing antivortex,
therefore yielding again the same result as the geodesic rule would predict:
no stable defects are formed. Again, the other antivortex resulting from the
phase wall decay escapes away from the bubbles.

Although we have not carried out simulations with phase walls of higher
winding, it is easy to convince oneself that the results will not change for
higher $|n|$ configurations. For instance, to reproduce with a higher
winding an initial situation analogue to the one in figure \ref{3bvortex}(a)
we would need a phase wall with $n=-3$. However, as this wall decays first
to $n=-2$ and then to $n=-1$ it nucleates into the region between the
bubbles first a vortex, then an antivortex. Obviously this pair annihilates
as it gets trapped between the bubbles, thus leaving us with the same
initial state as the one we have studied. The only difference would lie in
the number of defects that are expelled away from the bubbles.

\section{A model for how the geodesic rule works in phase transitions with
global U(1) symmetry}

Since the non-geodesic phase walls and their decay described in the sections
above were found to form in the region between the bubbles where the modulus
has an exponentially low value, the preceding considerations would seem to
be only an academic exercise. In this section we will however attempt to
give a physical picture of how the geodesic rule could work in models with
global symmetry.

We will start by assuming that the physical set up in which the process
takes place is that of a first order phase transition at finite temperature.
The oscillations in the value of the field around its false vacuum value due
to the finite temperature effects will yield a background thermal field.
These field fluctuations will each have a phase, and we will denote the mean
value of their modulus over space and time as $<\rho _{th}>$. They will also
have an average size (radius) and lifetime, $<\xi >$ and $<\tau >$. We will
also assume that $<\rho _{th}>$ is still well below the true vacuum value of
the field $\rho _{tv}$ so that the thermal fluctuations do not strongly
affect the field inside each bubble and the values of the phase of each of
them, $\theta _{1}$and $\theta _{2}$, remain well defined and largely
unaffected by the fluctuations. It is against this thermal background that
the bubble fields, $\phi _{1}$and $\phi _{2}$, will propagate. The $\rho
_{1,2}$ will obviously have the shape of the bubble profile, but any trace
of them will be lost soon after their value becomes lower than $<\rho _{th}>$
since beyond the bubble walls they decay exponentially with a typical decay
distance $\delta _{m}$. Therefore we will assume that beyond a few $\delta
_{m}$ from the point at which $\rho _{1,2}\sim <\rho _{th}>$ the thermal
background field will effectively erase any presence of the bubble fields.
Figure \ref{2bubbles} exemplifies this picture. The two bubbles will first
collide along the line that joins their centers. Roughly, this will happen
at about the time when their combined modulus at the point equidistant to
the centers $\rho _{b}$ --given by the ansatz (\ref{cmodulus}) evaluated at $%
x=0$-- reaches a value equal to the thermal fluctuation average $\rho
_{b}\sim <\rho _{th}>$, and the whole collision process will take place on a
region that we will characterize by a length $d$. Although $d$ will most
likely be of the order of a few $\delta _{m}$, we will leave it largely
unspecified for the time being. As we will see this will not have a big
impact in our conclusions as far as this simple argument is concerned. When
the collision takes place the phase has to decide which path will it follow
in the vacuum manifold in order to interpolate from $\theta _{1}$ to $\theta
_{2}$. Obviously the value of the phase of the thermal fluctuations in the
collision region will have an important influence on what configuration is
finally chosen, since by continuity the phase will be forced to interpolate
between $\theta _{1}$ and $\theta _{2}$ going through the value that it
takes at the collision region. After that, $\rho _{b}$ will rise above $%
<\rho _{th}>$ exponentially fast and as far the thermal effects are
concerned the configuration chosen by the phase at that point will remain
frozen thereafter. We will work in three spatial dimensions, and as
elsewhere in the paper we will take $\theta _{1}=0$ and $\theta _{2}=\pi /2$%
. This case is of particular interest because since $\Delta \theta $ will
always lie randomly between $0$ and $\pi $ (modulo $2\pi $), the average
phase difference between any two given bubbles will precisely be $\overline{%
\Delta \theta }=\pi /2$. We will examine two limiting cases.

As our first case let us examine a slow collision scenario in which the time
scale of the collision $t_{col}$ (i.e., the time needed by the bubbles to
expand their radii by an amount $d$\footnote{%
For bubbles expanding at the speed of light $t_{col}\sim d$, note however
that if the motion of the bubble walls is damped by interactions with the
external plasma (as the case seen in section I) $t_{col}$ can be
significantly larger than $d$.}) is much larger than the average lifetime of
the fluctuations $<\tau >$. If $t_{col}\gg <\tau >$, fluctuations appearing
between the bubbles when they are at a distance $d$ or less away from each
other will be energetically favored to have a phase that falls within the
geodesic path from $\theta _{1}$ to $\theta _{2}$. Under these conditions,
we will suppose that the probability $p_{n}$ of choosing any given
configuration with winding $n$ is weighted by a Boltzmann type of
distribution, $p_{n}\sim e^{-\beta E_{n}}$ where as usual $\beta =1/T$ (in
natural units, $k=1$), and $E_{n}$ is the energy corresponding to the given
configuration. Thus, the relative suppression factor for non-geodesic walls
of winding $n$ with respect to the geodesic configuration will be given by $%
e^{-\beta {\it \Delta }E}$, where ${\it \Delta }E=E_{n}-E_{0}$. Since both
solutions have the same background modulus for the field though, the
gradient of the modulus does not contribute to this energy difference, and
neither does the potential energy since the potential depends only on the
modulus. If we then assume that the collision region is small enough so that
the three dimensional energy density can be approximated by its one
dimensional limit, the total energy difference will simply be given by the
one dimensional integral times the cross sectional area of the collision
region, $d^{2}$. As was seen at the end of section II this approximation
will be valid for as long as $\left( d/x_{0}\right) ^{2}\ll 1$, with $x_{0}$
being the distance to the bubble centers, which will be justified in our
case. Thus, as an order of magnitude estimate we will have 
\begin{eqnarray}
{\it \Delta }E\sim d^{2}\int_{-\infty }^{+\infty }\rho _{b}^{2}\left(
\partial \theta _{n}^{2}-\partial \theta _{0}^{2}\right) dx\sim  \nonumber \\
<\rho _{th}>^{2}d^{2}\int_{-\infty }^{+\infty }\left( \partial \theta
_{n}^{2}-\partial \theta _{0}^{2}\right) dx,  \label{energyint}
\end{eqnarray}
since the phase gradients are exponentially close to zero outside of the
phase wall, where at the moment in which the collision takes place $\rho
_{b}\sim <\rho _{th}>$. We can now use the well known results for the
sine-Gordon kinks to get 
\begin{equation}
{\it \Delta }E\sim \frac{<\rho _{th}>^{2}}{\delta _{m}}d^{2}\left[
(4n+1)^{2}-1\right] .  \label{energydif}
\end{equation}
This is the key expression that will tell us how suppressed the non-geodesic
configurations will be. We can see how phase walls with high winding get
quickly suppressed, as one would have expected. As expected too, formation
of non-geodesic walls is also strongly suppressed for sufficiently high
values of $<\rho _{th}>$. This agrees with the standard argument which
states that at least for transitions with global symmetry the geodesic rule
must hold since it yields the least energetic configurations. Note however
how, for sufficiently low values of $<\rho _{th}>$ this argument doesn't
hold any more and production of non-geodesic walls could take place, in
particular those with low winding. The underlying physical reason is simply
that the energy density of the phase wall is given by $\rho ^{2}\partial
\theta ^{2}$, therefore, if the wall is placed in a region were $\rho ^{2}$
is small the energy difference between the different configurations will
also be small. Once the phase has chosen an interpolating path, the modulus
of the bubble field will quickly rise against the thermal background, and
the influence of the thermal fluctuations will become negligible as stated
above. The non-geodesic walls will subsequently decay, the pairs of defects
generated by their decay being expelled into the thermal background. It
would seem though that in this case we can expect to form non-geodesic phase
walls only for small values of $<\rho _{th}>$.

As our second limit let us examine the fast collision scenario and suppose
that $t_{col}\ll <\tau >$. In this case the last fluctuation on the scale of
the collision region appeared well before the bubbles were close to it. Let
us then make the naive assumption that, largely unaffected by the bubbles,
fluctuations in the collision region pick up phase values $\theta _{p}$ at
random from $0$ to $2\pi $ at the time at which they nucleate, with all
values having the same probability. Since by definition the geodesic path
from $\theta _{1}$ to $\theta _{2}$ is the shortest one on the vacuum
manifold, the field fluctuations in the collision region are more likely to
have a phase that falls {\it out} of the geodesic path than inside of it.
If, whatever value chosen for $\theta _{p}$, the paths travelled by the
phase to interpolate from $\theta _{1}$ to $\theta _{p}$ and then from $%
\theta _{p}$ to $\theta _{2}$ are geodesic, then it is easy to see that any
value of $\theta _{p}$ between $\pi $ and $3\pi /2$ will lead to the
formation of a $n=-1$ phase wall. That is, the probability $p(ng)$ of
forming a non-geodesic phase wall in this case will be simply 
\begin{equation}
p(ng)=\frac{\pi /2}{2\pi }=\frac{1}{4}.  \label{png}
\end{equation}
Note that this is an estimate from below, since phase walls with winding
other than $-1$ are not being considered at all in this simple picture.
(They would correspond to non-geodesic interpolations between the bubble
phases and $\theta _{p}$). Note also how in this limit the fact that for any
two bubbles $\overline{\Delta \theta }=$ $\pi /2$, as pointed above, also
implies that the average probability of forming a non-geodesic wall between
any two bubbles will be $0.25$, since the fact that $0\leq \Delta \theta
\leq \pi $ with all values having equal probability automatically implies
that $0\leq p(ng)<0.5$ (we exclude the case $\Delta \theta =\pi $ since in
that case any of the two paths is geodesic), and that $\overline{p(ng)}=0.25$%
. As before however, any nongeodesic phase wall will quickly decay into the
geodesic configuration, expelling vortices and antivortices into the
background.

The simple argument we have just presented would seem to indicate then that
in the case of slow collisions the geodesic rule is established mostly by
the thermal suppression of non-geodesic configurations, while for fast
collisions the rule holds because non-geodesic walls are unstable and decay
into the geodesic one. Since in both limits the geodesic rule holds, one
would also expect this to be the case in the (more physical) intermediate
regime in which $t_{col}\lesssim <\tau >$.

\section{Conclusions}

Although the geodesic rule is known to hold for first order phase
transitions that break a global U(1) symmetry, a dynamical account as to how
the field settles into the geodesic configuration was so far missing in the
literature. In this paper we have attempted to give precisely such an
account, in the hope that some of the findings and the methods here
developed could shed some light on the more difficult questions concerning
the validity of the rule in the Abelian-Higgs model. Using a family of
analytical solutions that represent the phase as it interpolates between its
values inside two different bubbles, i.e., phase walls, we have found both
the solution corresponding to the geodesic path as well as infinitely many
more which wind around the vacuum manifold $n$ times before reaching the
desired value. After finding the expression for the phase walls in 2+1
dimensions, we turned our attention to the question of the stability of the
non-geodesic configurations. Our analysis shows that, as one would expect,
non-geodesic phase walls are unstable, decaying into the geodesic
configuration in times of the order of $\delta _{m}$.

The decay of the non-geodesic states takes place via the nucleation of
vortex-antivortex pairs that quickly move away from each other, and the
simple rule that seems to be the result from our study is that a phase wall
of winding $n$ will produce precisely $|n|$ pairs of defects in its decay.
Although this production of defect pairs does not alter the geodesic rule
standard prediction of stable defect\footnote{%
i.e., defects surrounded by regions of true vacuum} nucleation in three
bubbles collisions, it is perhaps worth pointing out that for first order
phase transitions with supercooling (where $<\rho _{th}>$ will be low and $%
<\tau >$ large) a significant number of bubble collisions may potentially
result in the production of non-geodesic phase walls and their subsequent
decay. Thus, it would seem that for these transitions one could think of the
thermal background as of a region in which a gas of defects and antidefects
exist. Since the average modulus $<\rho _{th}>$ of this thermal field will
be low in supercooled transitions, the origin of these defects could in
general be due to both thermal fluctuations of the field and to their
production in phase walls decay. For as long as the second mechanism is not
dominant, one could think of this gas as being roughly in equilibrium,
since, in a large volume thermal fluctuations would be creating and
annihilating defects and anti-defects with roughly the same probability. If
production via phase wall decay becomes dominant though, the density of
defects in the thermal background will tend to grow as the transition
proceeds. A number of questions are raised by this scenario. For instance,
what will be the final fate of all these defects as the transition reaches
its end?. Will they all annihilate one another or will a significant
fraction survive the transition?. Although in a large volume the average
number of vortices will be equal to that of antivortices, in a single
smaller volume vortices and antivortices need not exactly cancel each other
out. What will happen if a three bubble collision traps a net winding of the
phase in the basin between the bubbles?. Also, could the presence of this
defect gas in the thermal background have an impact on models in which
baryogenesis is induced by topological defects?.

As far as the connection with the Abelian-Higgs model is concerned, some of
the results that we have found could be of interest. The motivation for our
study lies partly in the hope that one could export to the gauge case the
same type of techniques that we have developed to study the global model.
This could be done by defining some gauge invariant form of the phase
difference (i.e., {\it a la} Kibble-Vilenkin for instance \cite{geodesic},
where $\Delta \theta =\int dx^{k}{\it D}_{k}\theta $, with ${\it D}_{\mu
}\theta =\partial _{\mu }\theta +eA_{\mu }$, and $e$, $A_{\mu }$ being the
gauge coupling constant and gauge field respectively) and studying its
equations of motion. It is also worth pointing out the close resemblance
that some of our results bear to those of \cite{CS}. In this work, the
authors found out that both in the case of the Abelian-Higgs model and the $%
\sin \theta _{W}=0$ limit of the Weinberg-Salam model there may be enough
energy released from the collision of the bubble walls as to drive the
modulus of the Higgs away from its vacuum expectation value and make it
oscillate around zero. Extra regions of symmetry restoration are thus
produced, and the field oscillations as they decay lead to the formation of
multiple windings. In their simulations, they found that the validity of the
geodesic rule seemed to depend not only on the gauge coupling, but also in
the initial bubble separation and the phase difference between the bubbles.
Although in our model the energy needed to take the phase wall away from the
geodesic configuration is provided by the background thermal fluctuations,
rather than wall collision, note that our two dimensional phase walls also
show a dependence both on the distance between the bubble centers and on the
initial phase difference. (Although only for $\Delta \theta =\pi /2$ do we
have an analytical solution, solutions for other values of $\Delta \theta $
are in principle obtainable numerically). Furthermore, high winding phase
walls decay via field fluctuations around the false vacuum value, nucleating
multiple defects as they decay.

\section{Acknowledgments}

This project was greatly enriched by discussions with Alex Vilenkin, Pedro
Gonzalez\_Diaz and Guillermo Mena, to whom I wish to express my gratitude.
Special thanks are also due to Alejandra Melfo, for allowing me to use and
enrich the code that she first developed.

\newpage

\begin{figure}[tbp]
\caption{Lines of constant $\protect\theta $ for a geodesic phase wall going
from $\protect\theta =0$ to $\protect\theta =\protect\pi /2$. The lines are
labelled in units of $\Delta \protect\theta $. Thus, a label of $0.5$
implies a value of the phase of $\protect\pi /4$, and so on. The bubble
centers are placed at $x_{0}=\pm 10$, and we have taken $\protect\delta %
_{m}=1$. A lower value of $|x_{0}|$ would make the contour lines show more
curvature for the same axis scale, whereas for larger values the wall would
look flatter on this scale. }
\label{2dpwall}
\end{figure}
\begin{figure}[tbp]
\caption{The decay of a $n=-1$ phase wall is shown in polar coordinates in
field space. Distance to te origin gives the modulus of the field, while the
angle corresponds to the field phase. In (a) the initial configuration at $%
t=0$ is shown. (b) corresponds to $t=0.4$, (c) to $t=0.6$, and finally (d),
in which we can see that the geodesic wall is the final state, corresponds
to $t=2$. Time is measured in units of the bubble wall thickness $\protect%
\delta _{m}$. }
\label{1dnm1}
\end{figure}
\begin{figure}[tbp]
\caption{We see here the decay of the $n=-2$ phase wall into a configuration
equivalent to a $n=-1$ wall. As before, the initial $t=0$ configuration is
shown in (a). Subsequent images correspond to: (b) $t=0.1$, (c) $t=0.15$,
and (d) $t=0.2$ .}
\label{1dnm2}
\end{figure}
\begin{figure}[tbp]
\caption{The shrinkage of the extra $2\protect\pi $ loop is shown here in
greater detail. Images correspond to: (a) $t=0$, (b) $t=0.11$, (c) $t=0.13$
by which time the field has already crossed the origin and we no longer have
a $2\protect\pi $ phase change around the loop, (d) $t=0.14$ where we see
how the loop has been stretched into a cusp, (e) $t=0.16$ where the cusp has
divided into two sharp kinks that move away from each other, and finally (f) 
$t=0.18$ . }
\label{1dnm2b}
\end{figure}
\begin{figure}[tbp]
\caption{A $n=-1$ two dimensional phase wall decays into the geodesic
configuration. Solid lines indicate contour lines of the field modulus,
while the phase is given by the angle between the arrows and the $x$ axis.
The length of the arrows is proportional to the value of the mudulus at the
point they are drawn, except in the central region occupied by the phase
wall, wher they have been magnified to a common length. The initial
configuration at $t=0$ is shown in (a). In (b) $t=0.3$, and we see the
modulus of the field quickly decreasing in the central region. In (c) $%
t=0.304$, the field has crossed the origin in field space at the central
point, and its modulus is now growing there. As described in the text this
produces a vortex -antivortex pair, we can see the vortex somewhat above $%
y=1 $, and the antivortex at $y=-1$. In (d) $t=0.308,$ the defects keep
moving away from each other as the field increases its modulus in the
central region. }
\label{2dnm1}
\end{figure}
\begin{figure}[tbp]
\caption{A $2$-dimensional $n=-2$ phase wall decays into the $n=-1$%
configuration. The initial coniguration is shown in (a) at $t=0$. To see the
extra $2\protect\pi $ factor in the phase follow its change along a
horizontal line, the arrows turning always in a clockwise direction. In (b) $%
t=0.108$, the field is decreasing its value in the central region, in (c) $%
t=0.109$ and the vortex (around $y=-1$) and antivortex ($y=1$) have already
appeared. Note how between the defects the phase is now increasing across
the wall from almost $-\protect\pi $ to almost $-\protect\pi /2$,
subsequently decreasing again until it reaches $-3\protect\pi /2$. In (d), $%
t=0.111$, the defects keep moving away from each other, and the field
increases its modulus in the central region.}
\label{2dnm2}
\end{figure}
\begin{figure}[tbp]
\caption{a $n=-1$ phase wallbetween two bubbles with phases $5\protect\pi /6$
(left) and $\protect\pi /6$ (right) decays in the presence of a third bubble
with phase $3\protect\pi /2$. In (a) $t=0$, (b) $t=0.34$, (c) $t=0.384$, and
(d) $t=0.8$. The decay sends a vortex into the region between the three
bubbles that gets trapped there. The (upper) antivortex is expelled away
from the bubbles.}
\label{3bvortex}
\end{figure}
\begin{figure}[tbp]
\caption{A $n=-1$ wall decays in the presence of a third bubbe. The three
bubbles would not form a defect according to the geodesic rule. The net
result of the process agrees with that predicition. In (a) $t=0$, we can
already see how, as a consequence of the presence of the $n=-1$ wall we
already have an antivortex in the region between the bubble centers. The
wall subsequently decays, in (b) $t=0.34$, and (c) $t=0.372$, sending a
vortex into the interior region. This vortex and the antivortex annihilate
each other, leaving in (d) $t=0.8$ the final configuration predicted by the
geodesic rule.}
\label{3bnovortex}
\end{figure}
\begin{figure}[tbp]
\caption{Two bubbles bubbles about to collide expand against a fluctuating
thermal field background. Since the bubble fields decay exponentially fast
beyond the walls, any trace of them is quickly lost once they get below the
thermal field level. The thermal fluctuations are presumed not to be strong
enough to significantly alter the bubble fields inside the walls however.}
\label{2bubbles}
\end{figure}

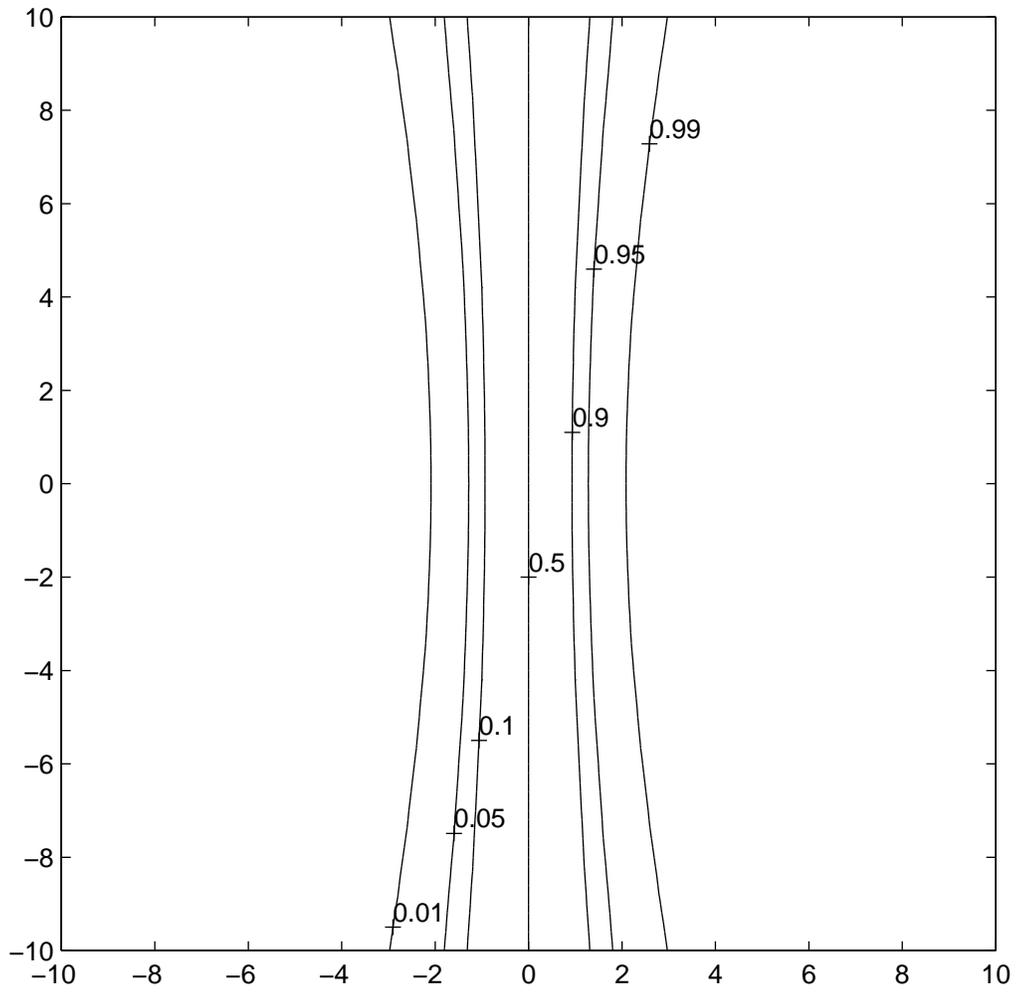
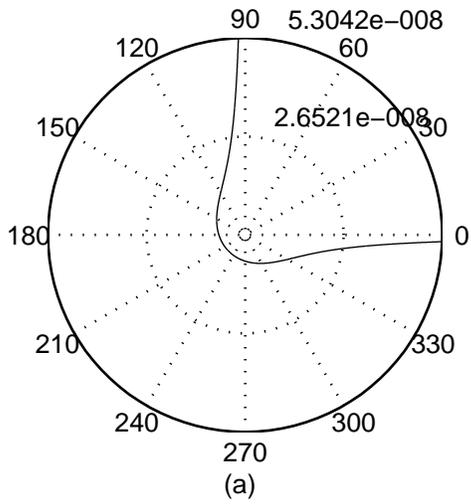
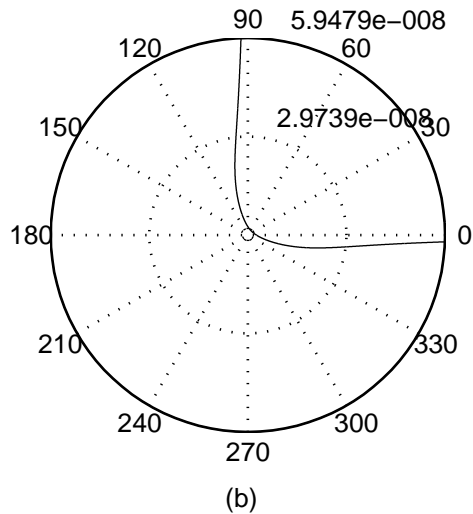
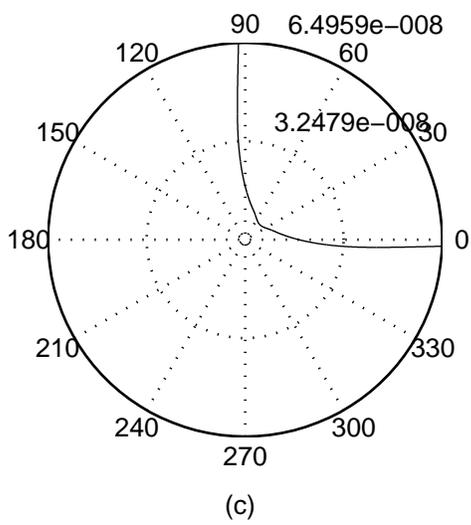
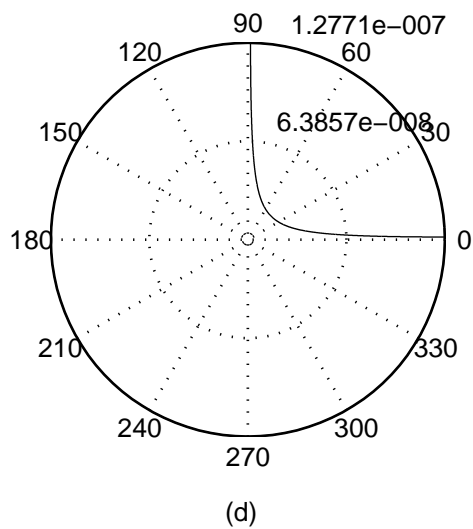
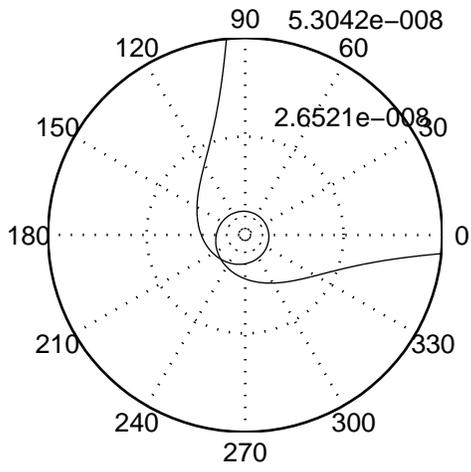
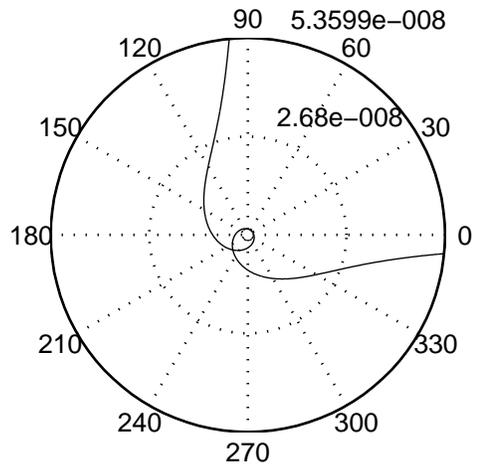
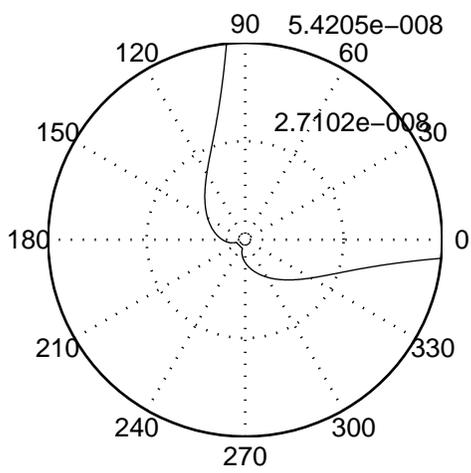
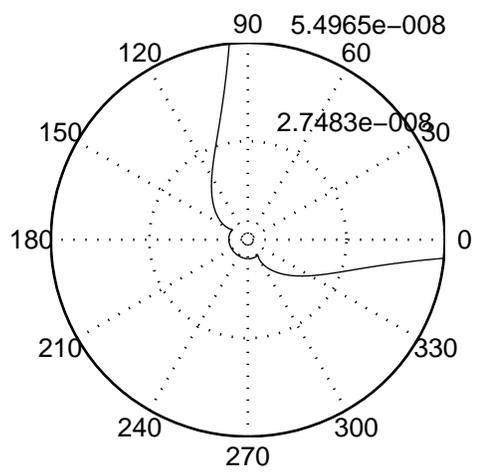
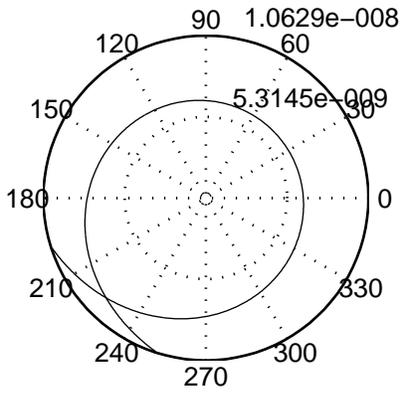
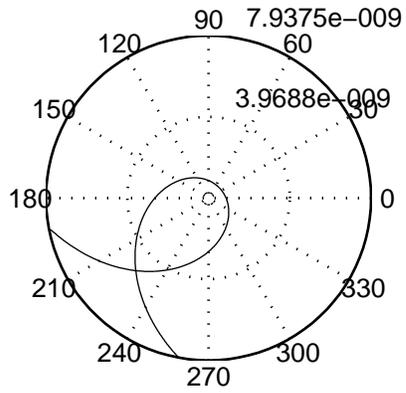
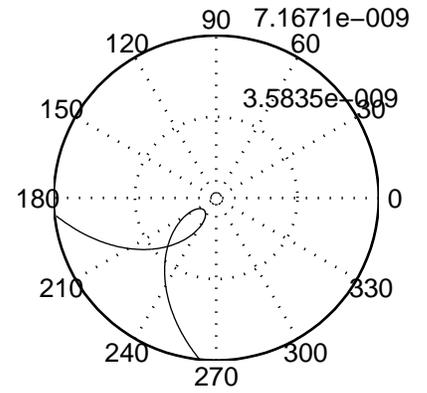
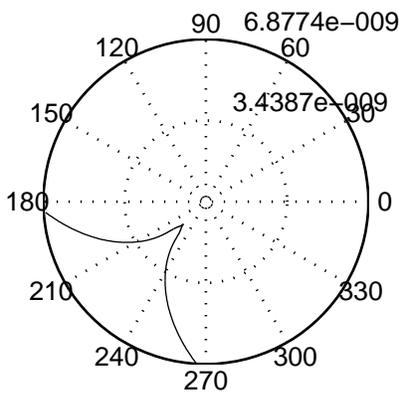
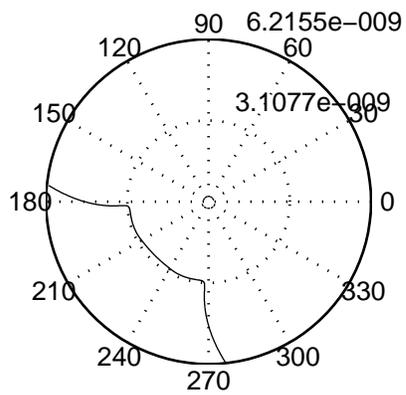
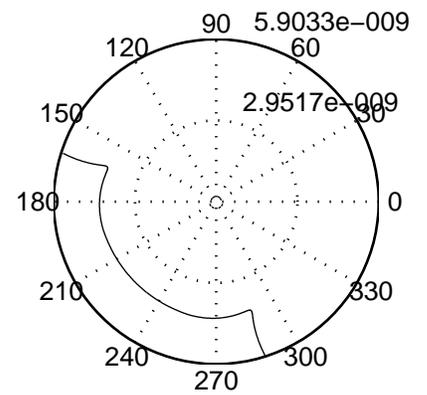
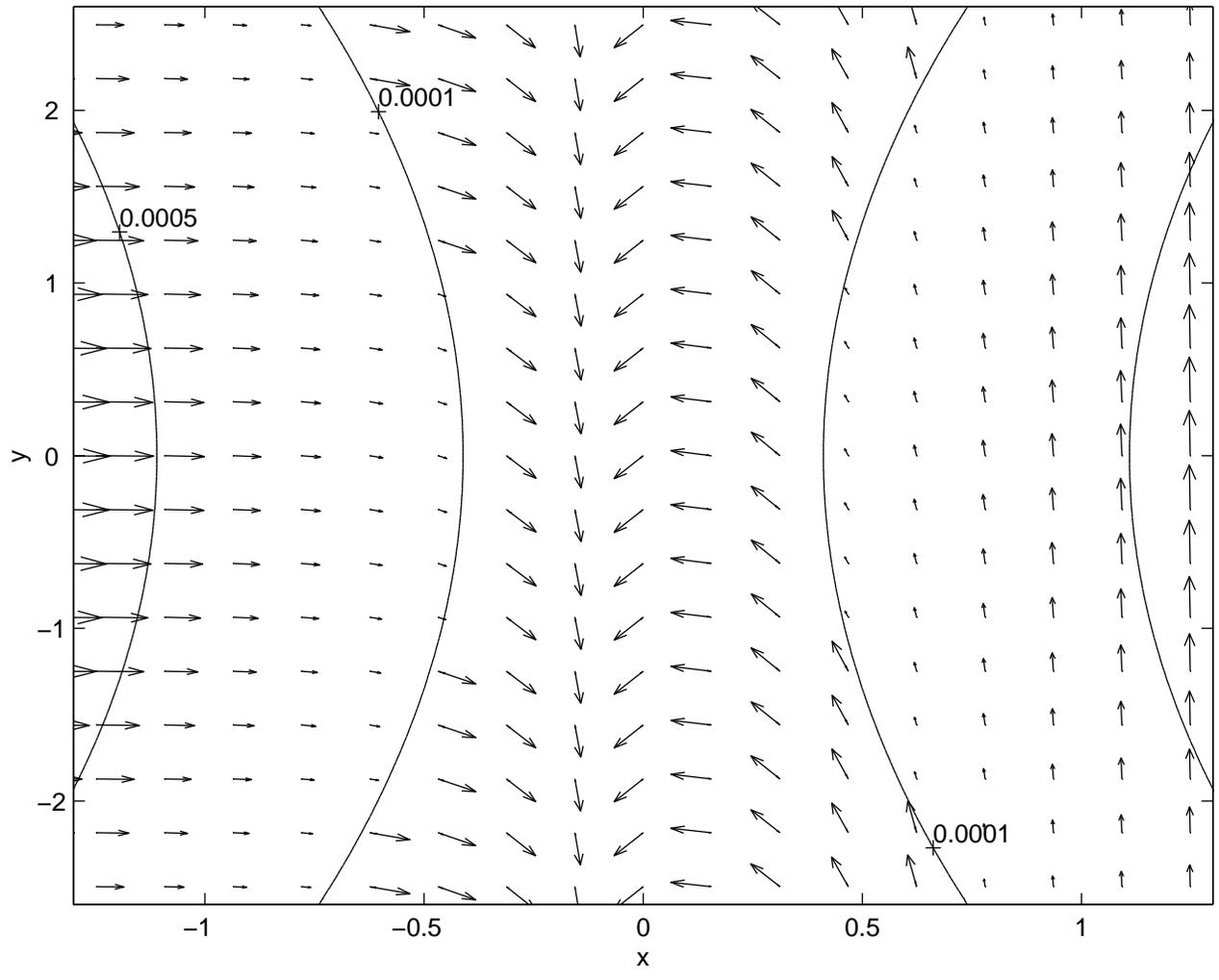
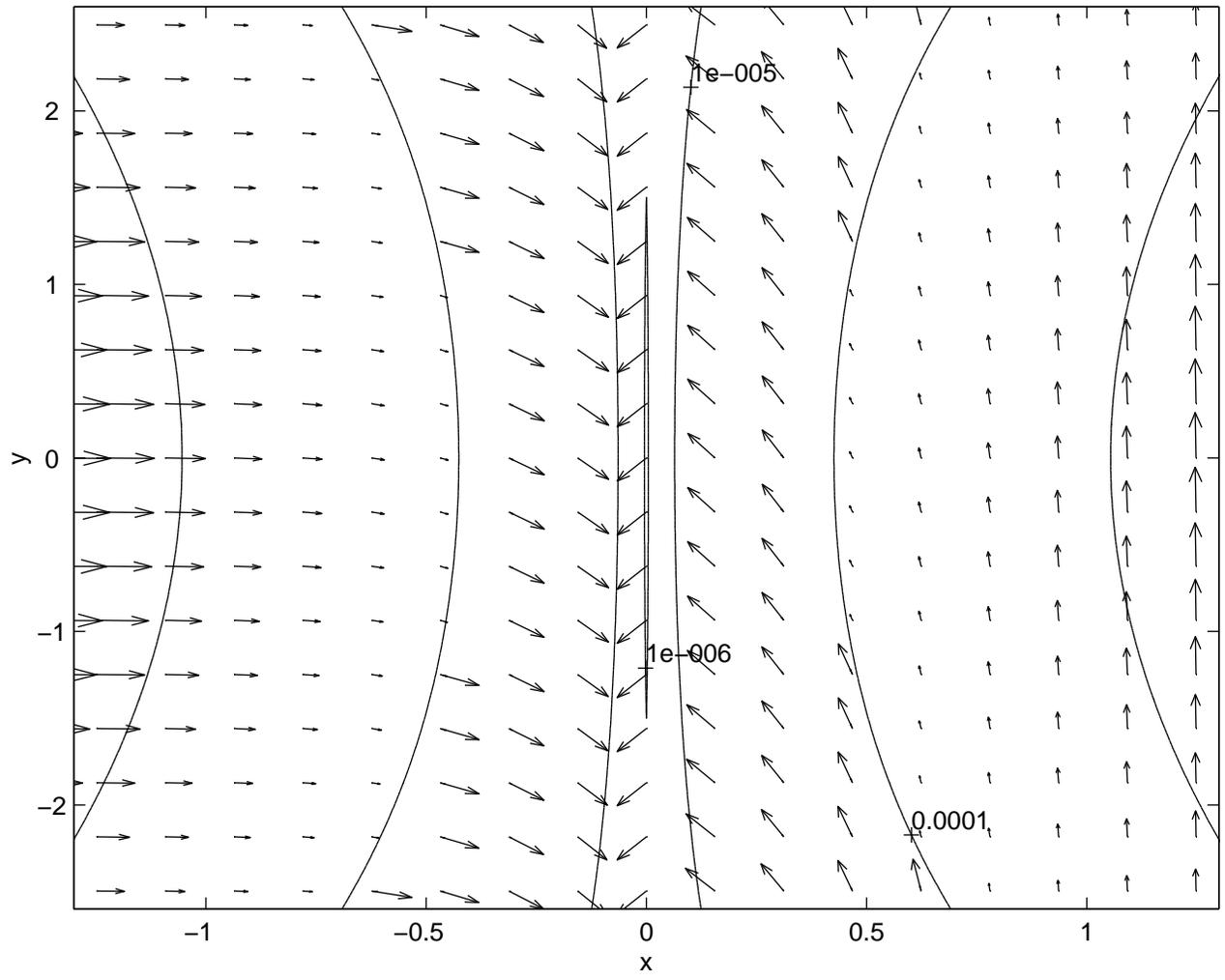
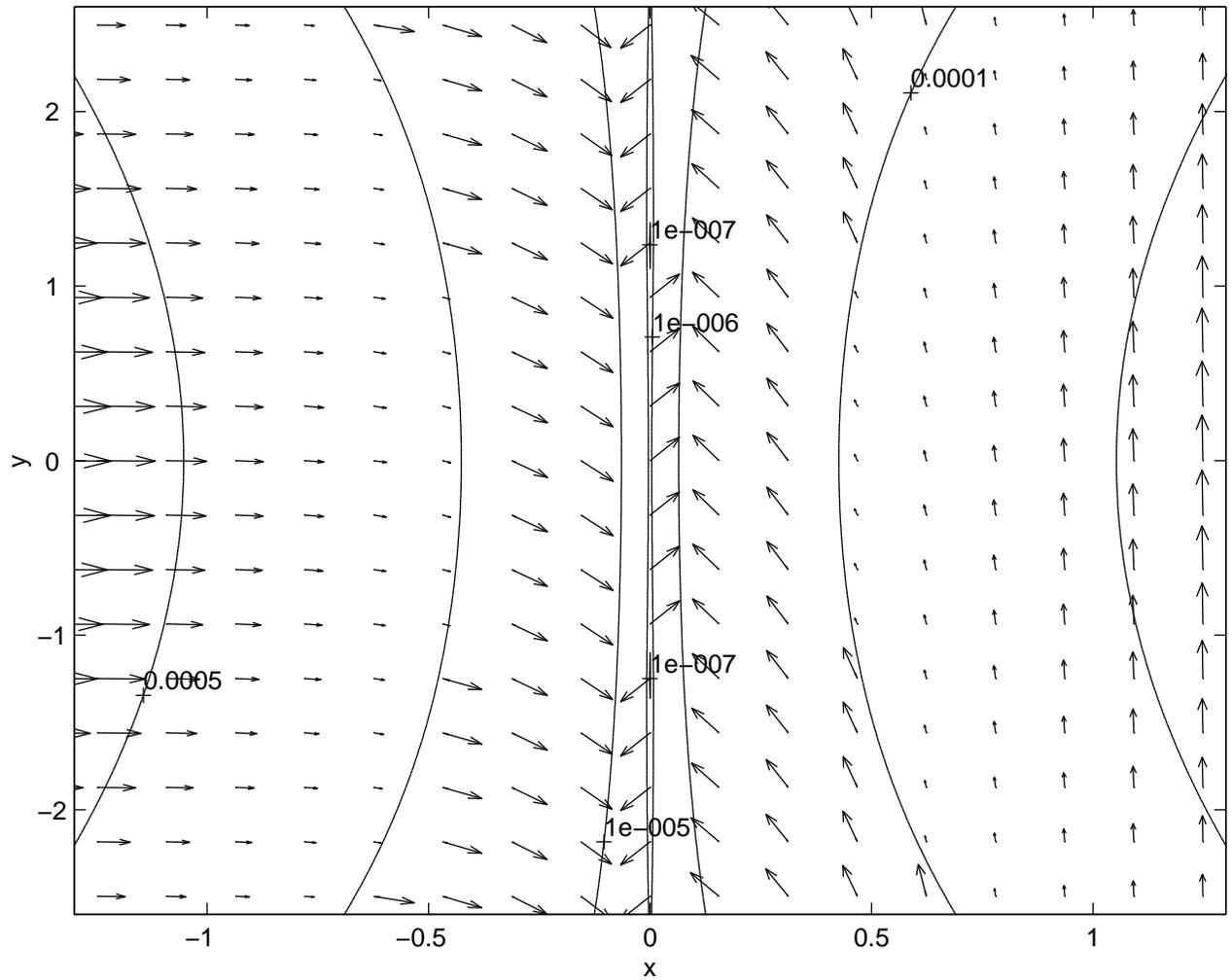
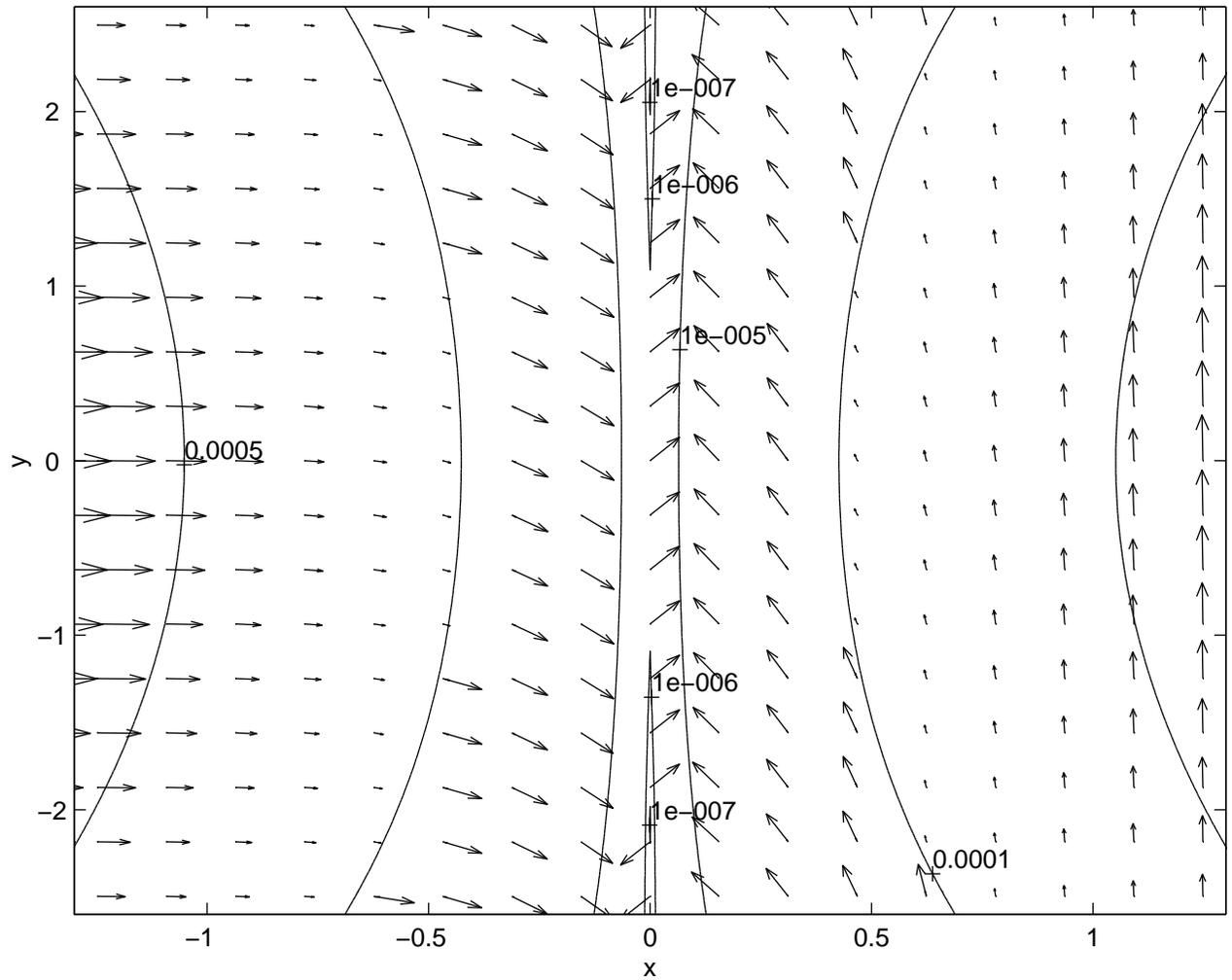
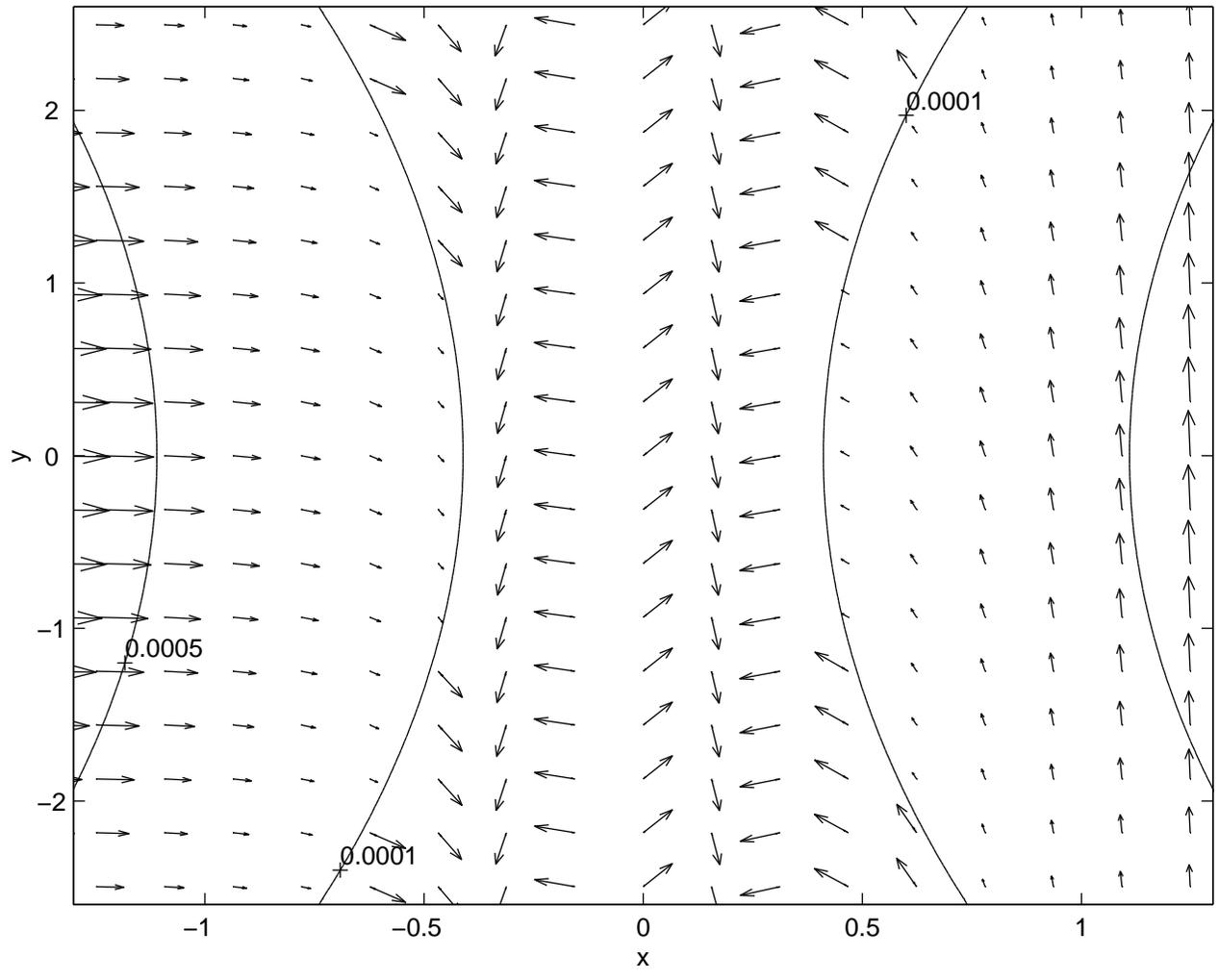
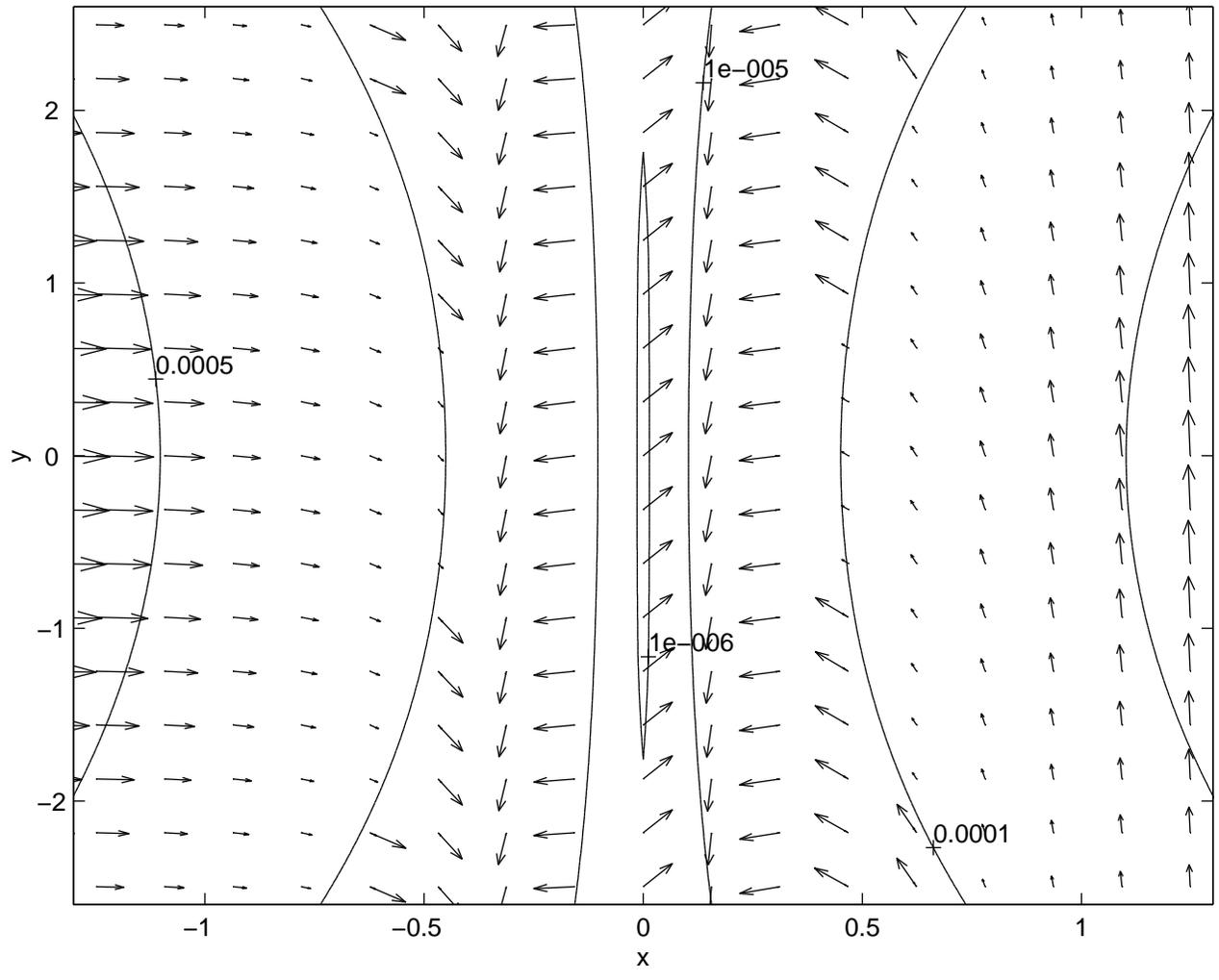
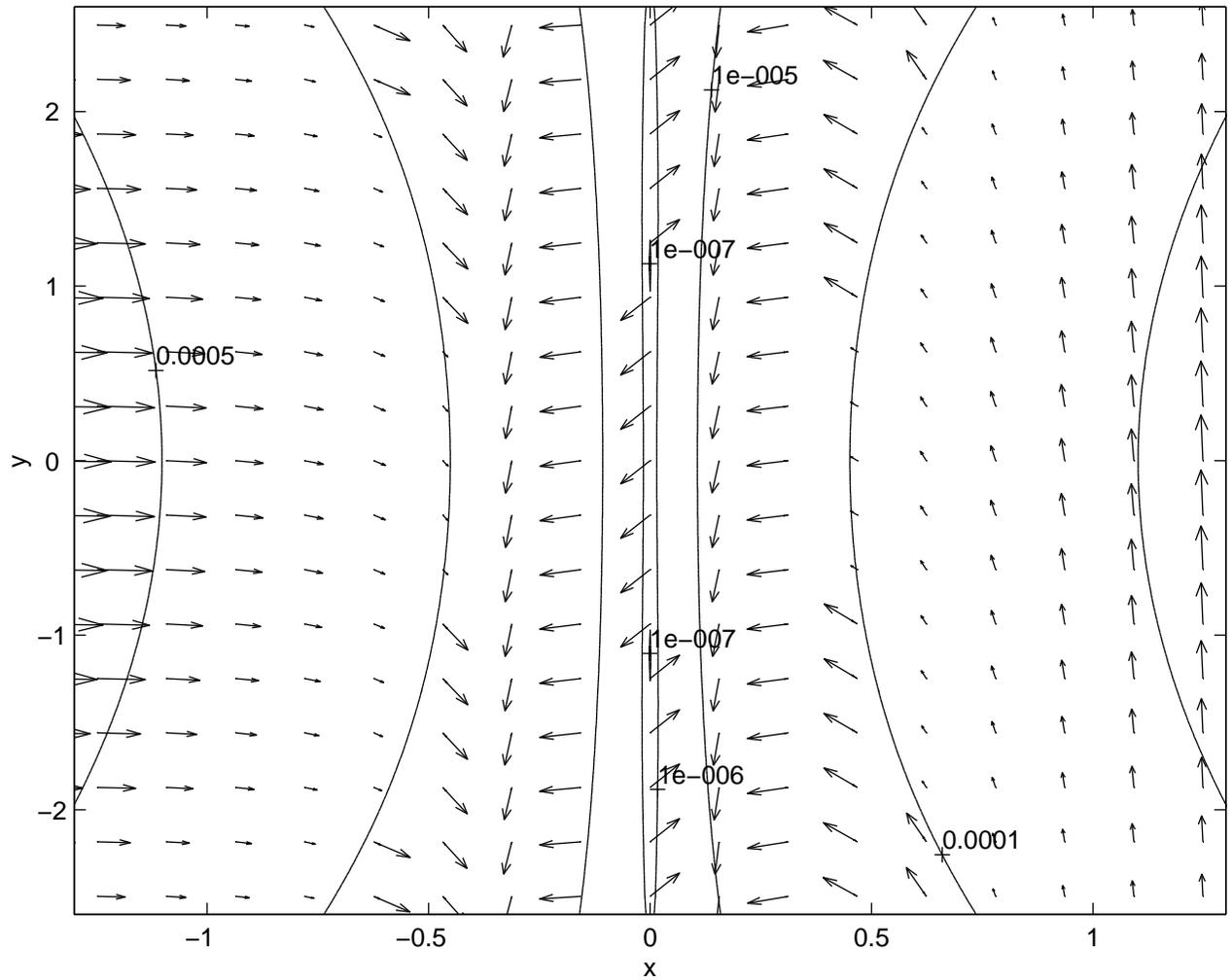
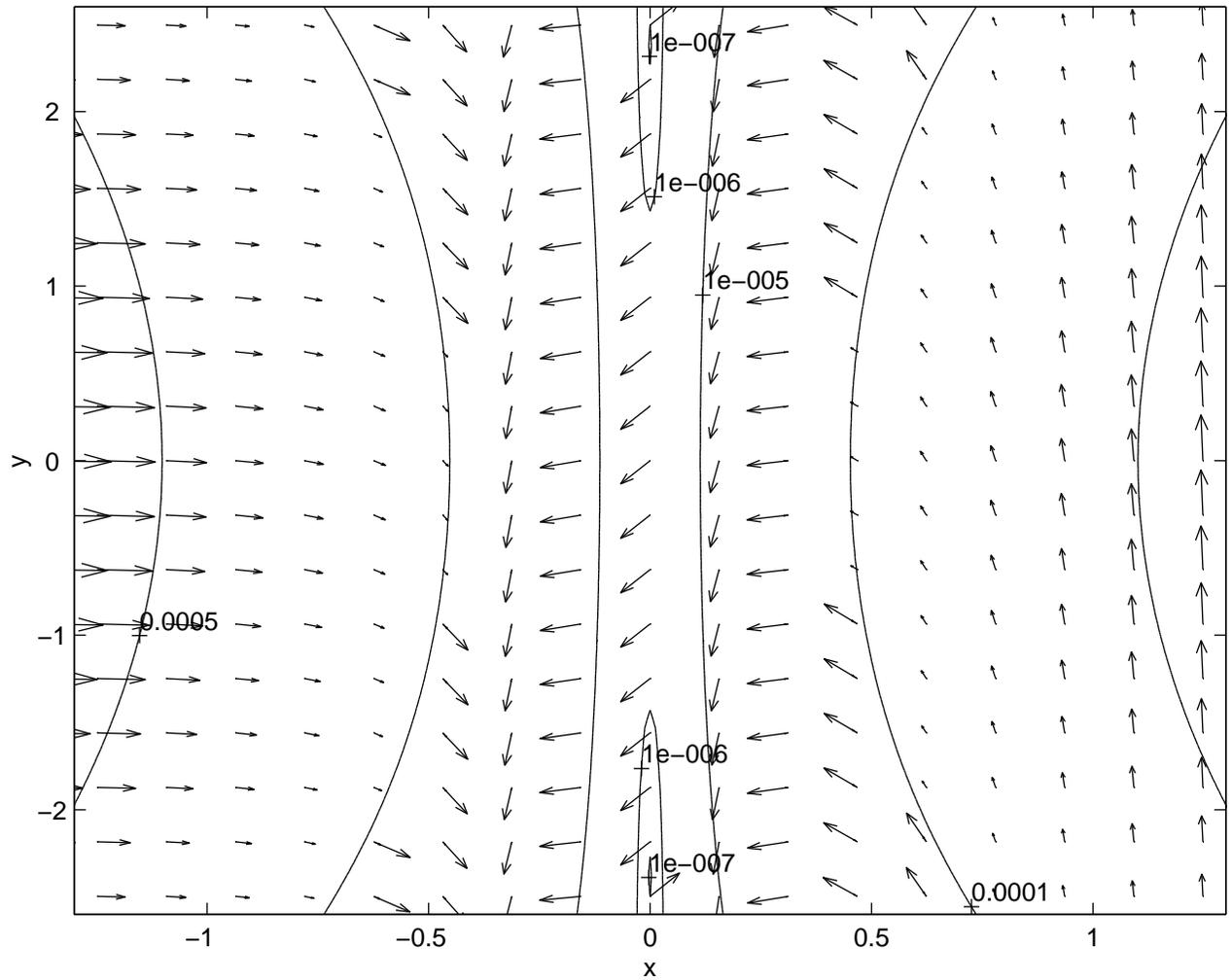
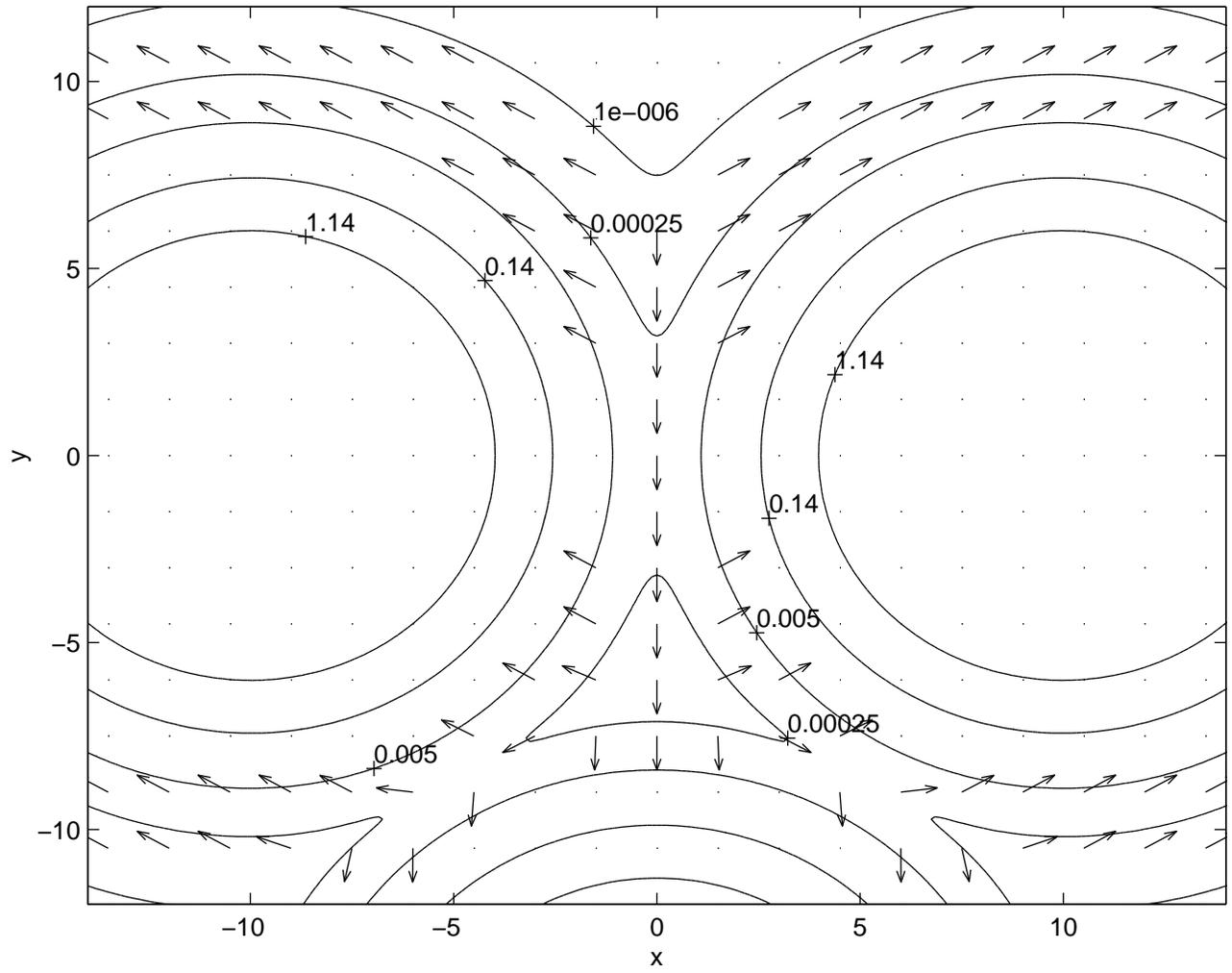
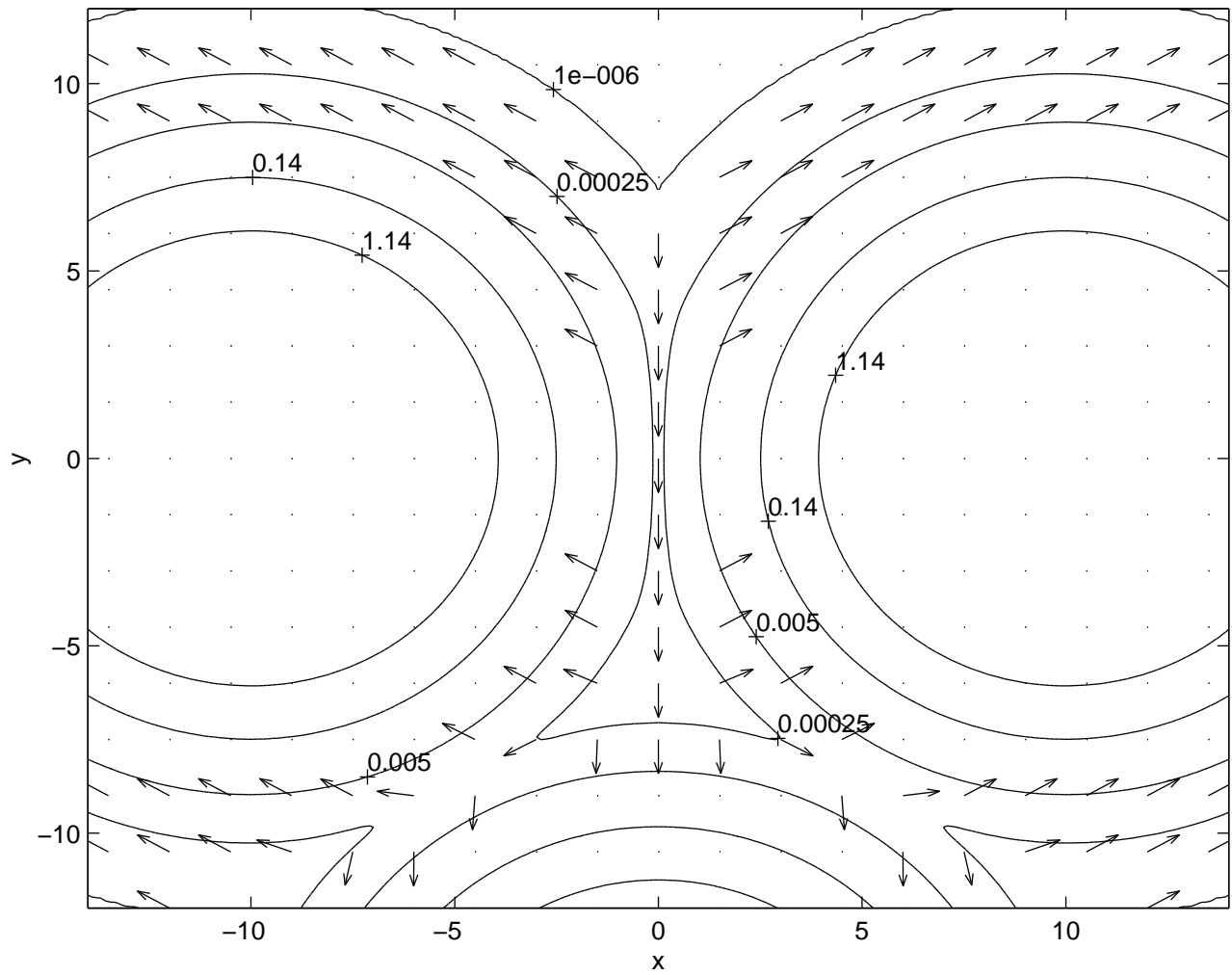
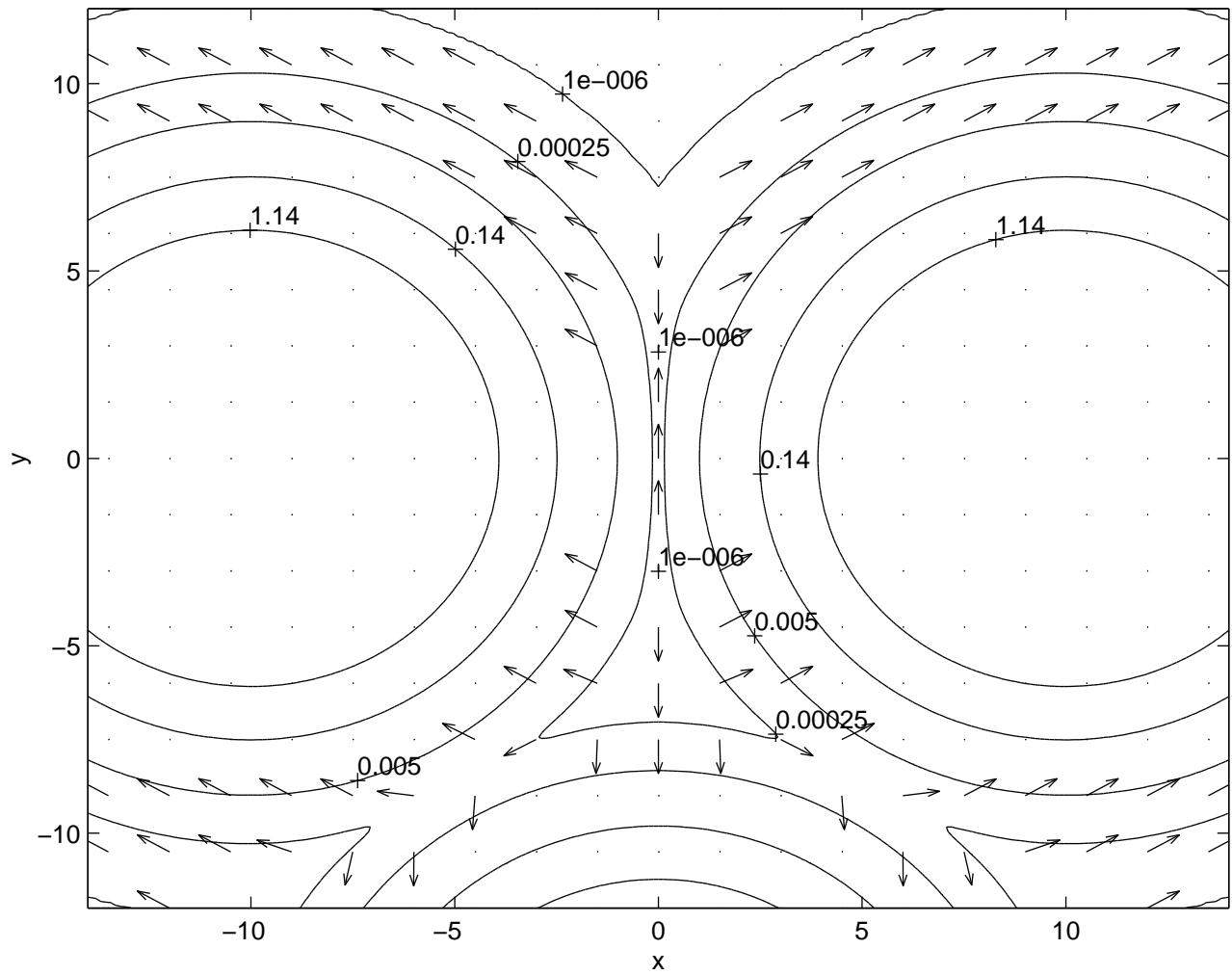
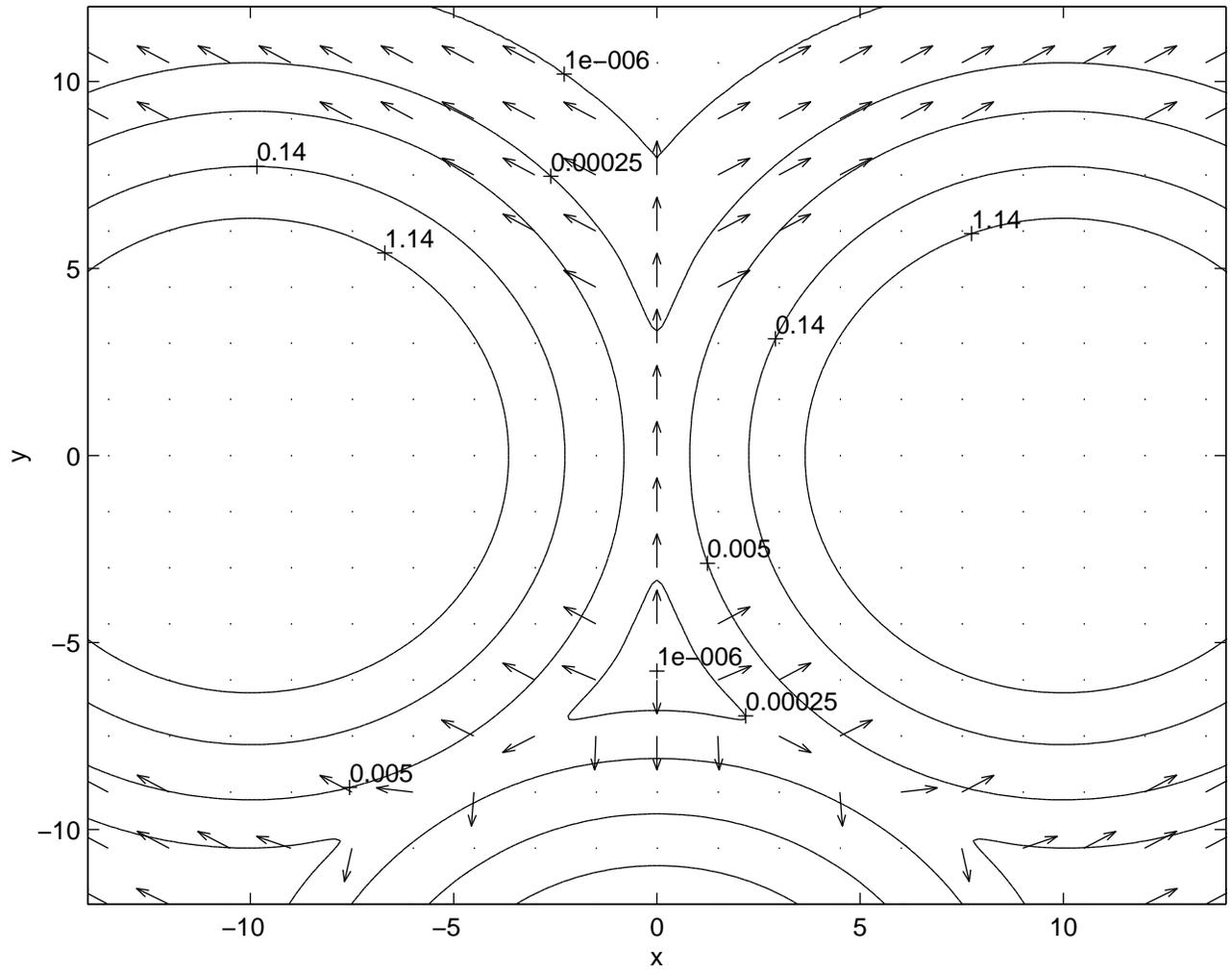
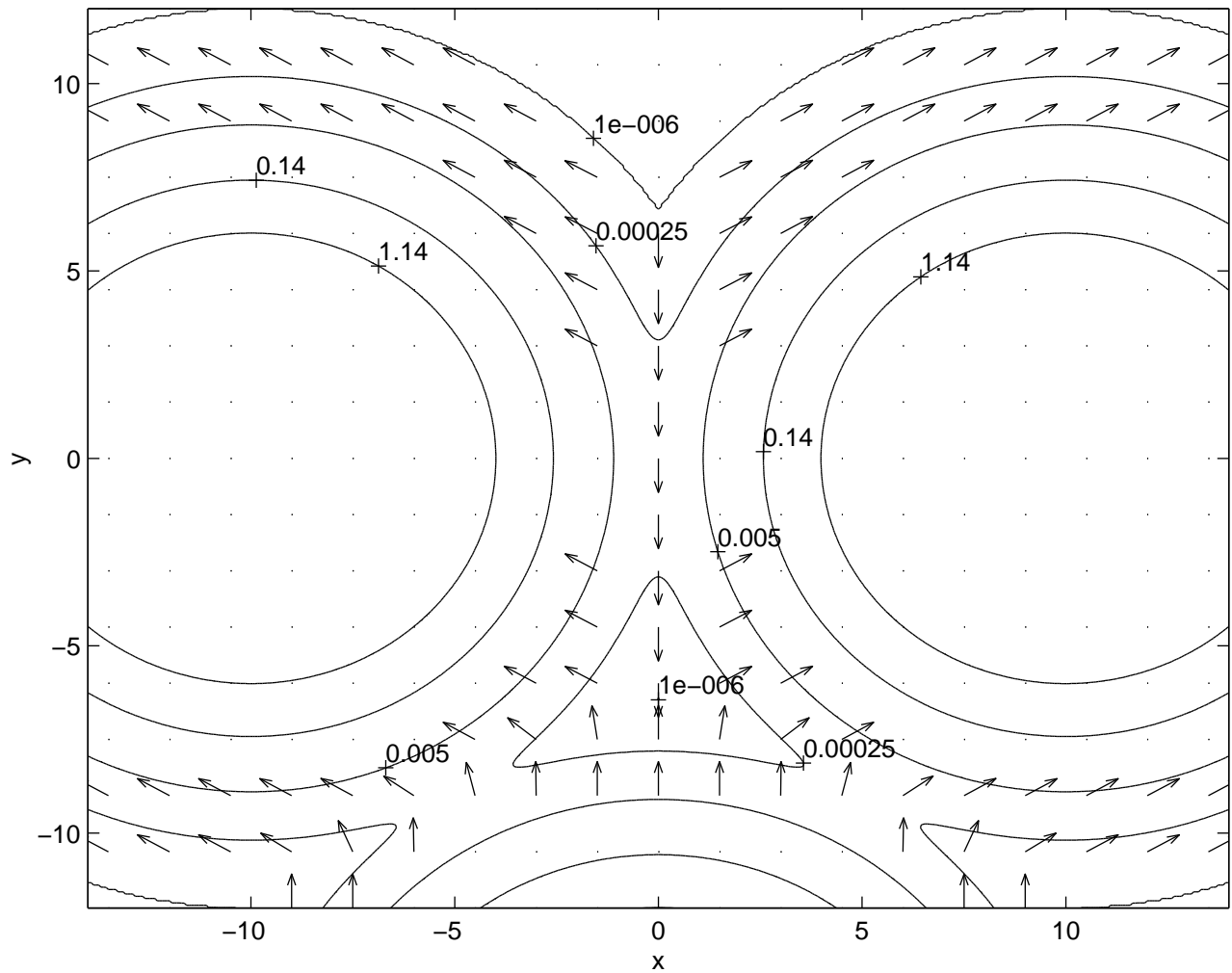
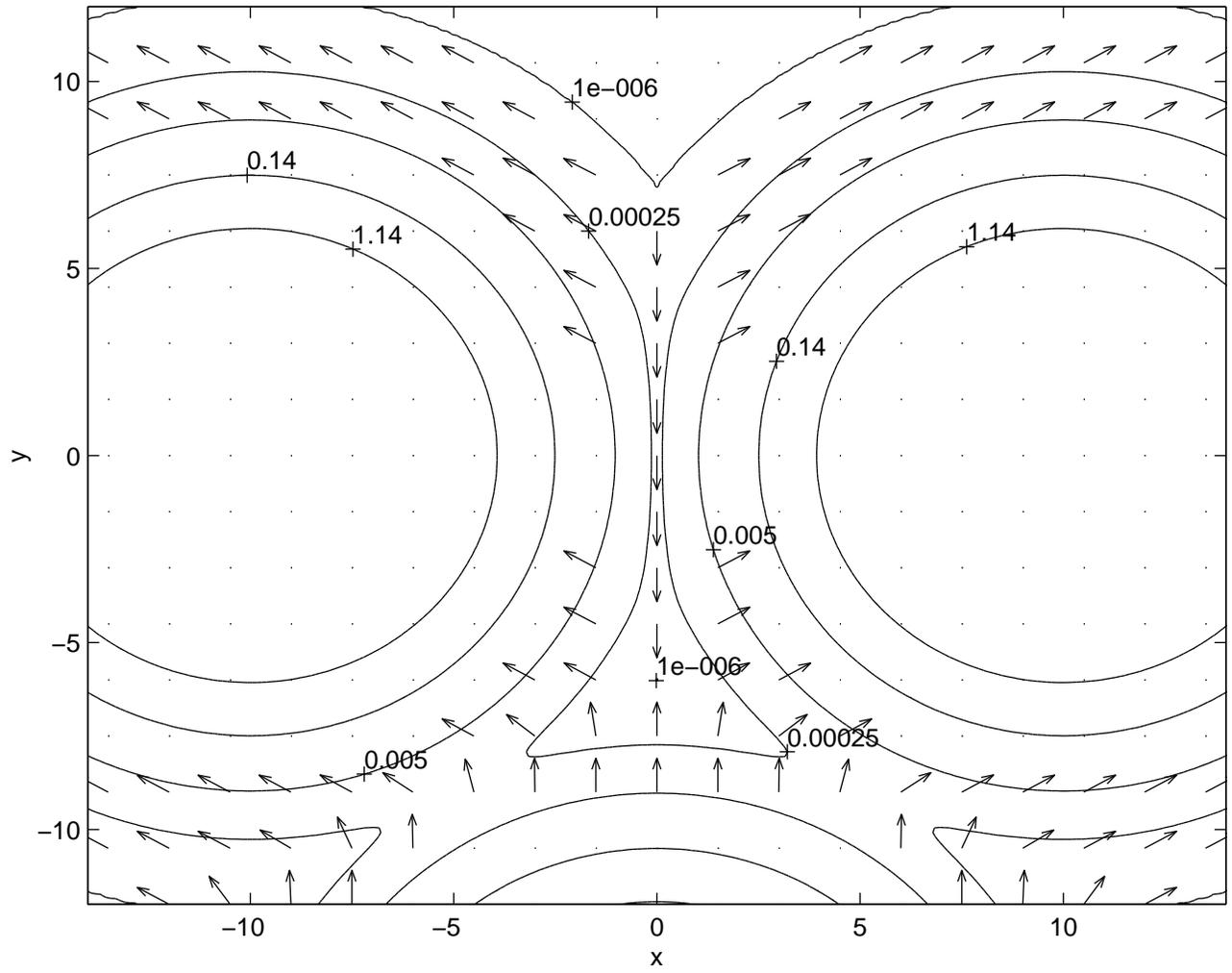
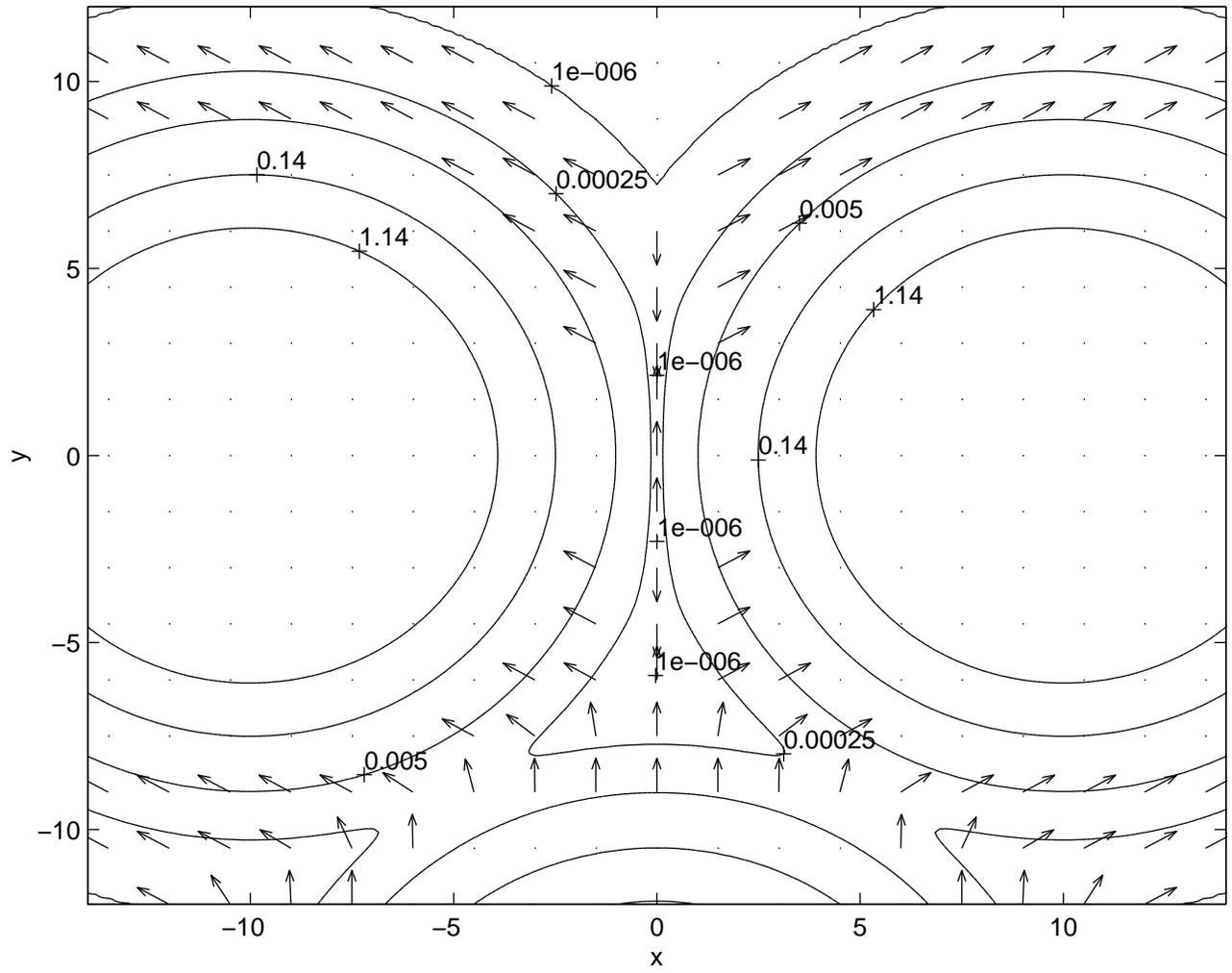
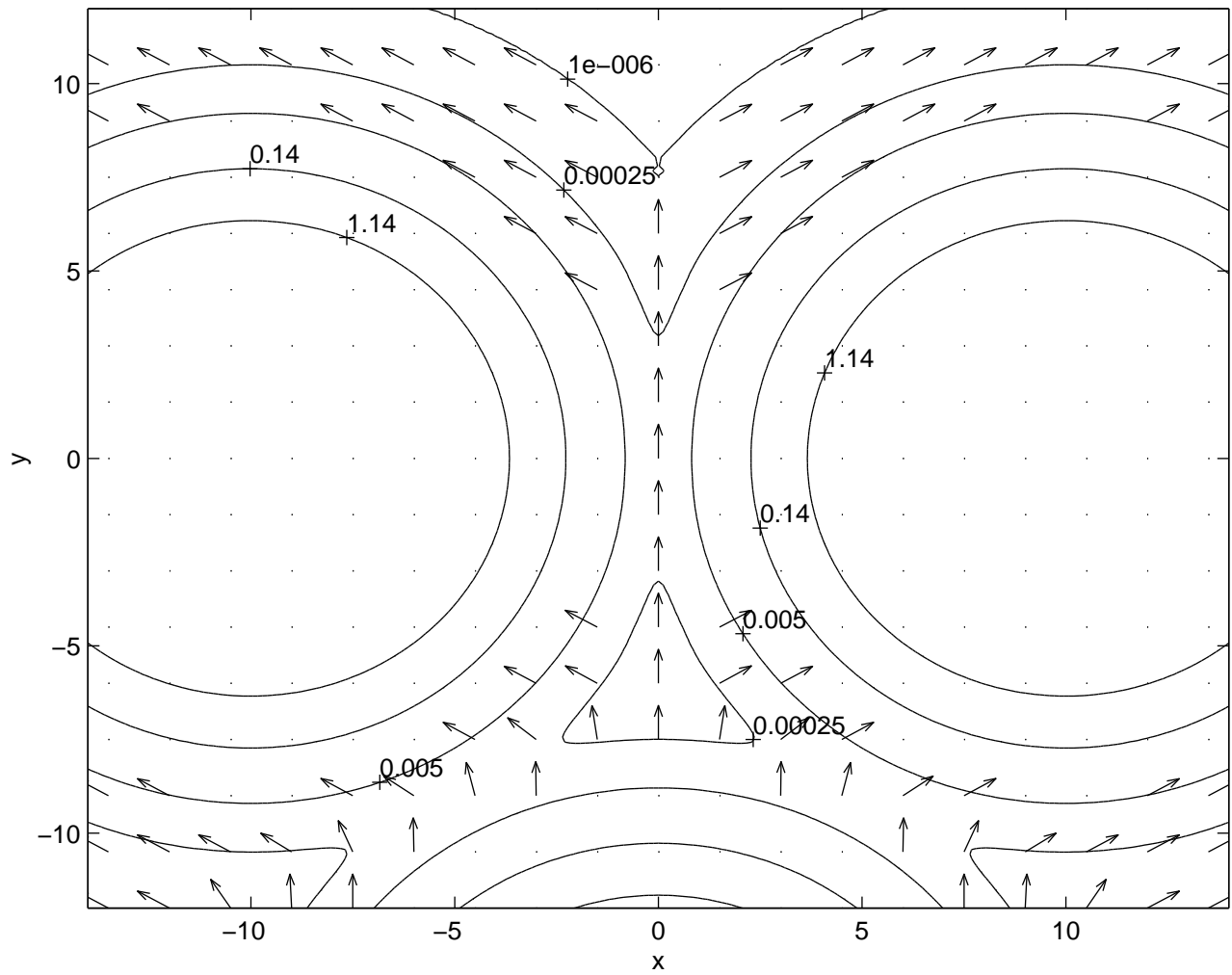
\end{document}